\documentclass{jfm}
\usepackage{graphicx}  
\usepackage{esvect}
\usepackage{epstopdf,epsfig}
\usepackage{textgreek}
\usepackage[scr=bickham]{mathalfa}
\usepackage{amsmath}
\numberwithin{equation}{section}
\usepackage{subfigure}\usepackage{float}\usepackage{epsfig}
\usepackage{amssymb}\usepackage{psfrag}\usepackage{natbib}
\usepackage{stmaryrd}
\usepackage{picture}
\usepackage{float}
\usepackage[dvipsnames]{xcolor}
\usepackage[T1]{fontenc}
\usepackage{multirow}
\usepackage{overpic}
\usepackage{tikz}
\usepackage{subcaption}\usepackage{float}\usepackage{epsfig}
\DeclareCaptionLabelFormat{new}{\textup{(}{\text{#2}}\textup{)}}
\usepackage{stmaryrd}
\usepackage{float}
\usepackage{bm}
\usepackage{multirow}
\usepackage{makecell}
\usepackage{ragged2e}

\newcommand{\be}{\begin{equation}}
\newcommand{\ee}{\end{equation}}
\newcommand{\bef}{\begin{figure}}
\newcommand{\ef}{\end{figure}}
\newcommand{\bear}{\begin{eqnarray}}
\newcommand{\ear}{\end{eqnarray}}
\newcommand{\barr}{\begin{array}}
\newcommand{\earr}{\end{array}}

\newcommand{\bs}{\boldsymbol}

\newcommand{\msb}{\mathbf}

\definecolor{bittersweet}{rgb}{1.0, 0.44, 0.37}
\definecolor{bluep}{rgb}{0.2, 0.2, 0.6}
\definecolor{bole}{rgb}{0.47, 0.27, 0.23}
\definecolor{cgreen}{rgb}{0.0, 0.42, 0.24}
\definecolor{dpurple}{rgb}{0.59, 0.44, 0.84}

\usepackage[dvipsnames]{xcolor}
\usepackage{ragged2e}
\usepackage{physics}
\usepackage[colorlinks=true,breaklinks=true,linkcolor=blue]{hyperref} 
\shortauthor{Arun Krishna B. J. and Ganesh Tamadapu}
\title{Nonspherical oscillations of an encapsulated magnetic microbubble }

\author{Arun Krishna B. J.\aff{} \and Ganesh Tamadapu\aff{}
  \corresp{\email{gt@iitm.ac.in}}}

\affiliation{
  Department of Applied Mechanics and Biomedical Engineering, Indian Institute of Technology Madras,
Chennai, 600036, India.}
\begin{document}
\maketitle
\begin{abstract}
This paper presents a model for nonspherical oscillations of a lipid-encapsulated bubble infused with magnetic particles, developed using membrane theory for thin, weakly magnetic membranes. The model assumes that only the applied magnetic field contributes significantly to the Maxwell stress, with the membrane under generalized plane stress. Axisymmetric deformations are analyzed under two external magnetic field configurations: symmetrically arranged current-carrying coils and symmetrically placed magnetic dipoles. The non-spherical oscillations are restricted to the linear regime, with the second mode dominating within the stable pressure range. The pressure–frequency stability region is computed, and its dependence on material properties and magnetic forcing is examined, and natural frequencies are estimated using a boundary-layer approximation. In the coil-field configuration, time-series analysis shows that the second mode attains significant amplitude relative to the bubble radius, increasing with both interfacial magnetic susceptibility and initial radius. Order-of-magnitude estimates indicate that pressure forcing dominates over magnetic forcing, implying a negligible effect of applied current on radial oscillations and stability. In contrast, for the dipole field, increasing dipole strength and reducing its distance from the bubble center significantly reduces the stability region. Simulations with time-varying current further show that its amplitude and frequency have minimal influence on the stability diagram but strongly affect the periodicity of mode shapes.
\end{abstract}

\maketitle
\section{Introduction}\label{s:1}
Encapsulated microbubbles have become essential in biomedical applications, serving as contrast agents for ultrasound imaging and as targeted drug delivery carriers. The development of innovative all-in-one drug delivery systems has garnered significant attention due to their multifunctionality, enabling the integration of therapeutic and imaging components, as well as targeting moieties, for simultaneous targeted therapy and imaging \citep{stride2009enhancement,sciallero2016magnetic,chertok2018circulating}. \citet{mulvana2012theoretical} in their work studied experimentally the oscillation of a bubble encapsulated with magnetic particles in the presence of a magnetic field. Adding magnetic nanoparticles to encapsulation have three key additional features making them attractive for biomedical applications: their ability to be visualised under MRI \citep{sun2008magnetic}, to be localised at a target site by means of a magnetic field gradient \citep{pankhurst2003applications} and to produce localised heating when exposed to an oscillating magnetic field for hyperthermia and/or to promote drug delivery \citep{duran2008magnetic,thiesen2008clinical}. \citet{latulippe2014dipole,7322277} used dipole fields to navigate contrast agents for targeted drug delivery, which provided  depth-independent saturation magnetization of the contrast agents and high gradient strengths, while
being adapted to whole-body interventions. Further, it has been reported in~\cite{duan2015controlled} that an increase in the number of magnetic nanoparticles loaded onto the micro-bubbles resulted in magnetic resonance image enhancement and increased the duration for which the ultrasound image could be observed. \citet{owen2015magnetic} analyzed the dynamics of magnetic bubbles, assuming them to be rigid, and studied the flow conditions necessary to retain these bubbles near the vessel wall in a Poiseuille flow under an applied magnetic field. However, designing multifunctional agents that meet specific diagnostic and therapeutic requirements remains a significant challenge. This is largely due to the limited understanding of how the integration of magnetic nanoparticles within the microbubble shells influences their mechanical properties and dynamic behavior in both ultrasound imaging and drug delivery systems. 
Addressing these complexities is crucial for the development of effective multifunctional agents. The challenge in theoretically modeling the oscillations of encapsulated magnetic microbubbles arises from the absence of magnetic monopoles, which makes the oscillations inherently axisymmetric. Non-spherical modes are not introduced as parametric instabilities of the spherical mode; rather, they are a direct result of the applied magnetic field, which excites these modes. The nonspherical behavior of gas bubbles is well established both theoretically and experimentally. \citep{hao1999effect,guedra2016experimental,guedra2018bubble,shaw2006translation, shaw2009stability,shaw2017nonspherical}. The numerical investigation by \citep{shaw2006translation} and \citep{guedra2016experimental} showed that the nonlinear mode coupling is responsible for the saturation of instabilities. 

There are relatively few studies focused on modeling the non-spherical oscillations of encapsulated bubbles. \cite{tsiglifis2011parametric} investigated the parametric stability and dynamic buckling of encapsulated bubbles, while \cite{liu2012surface} examined the surface instability of these bubbles and analyzed how membrane surface properties influence the natural frequencies of the bubble shell. They also showed that the parametric resonance of encapsulated bubbles is similar to that of gas bubbles, with membrane effects having no significant impact on stability. However, the above models are not completely valid as they consider only linear oscillations of shape modes. \cite{dash2024nonspherical} incorporated nonlinear mode coupling and interface energy into their model, demonstrating that non-linearity also saturates instability in the case of encapsulated bubbles.

Till now, a comprehensive model for encapsulated magnetic microbubbles remains elusive. \cite{zhao2022nonlinear} modeled encapsulated bubbles undergoing radial oscillations in the presence of a magnetic field, while \cite{du2024local} extended this by incorporating translational motion along the $x$ and $y$ directions in addition to radial oscillations. However, these models are incomplete, as purely radial oscillations are unrealistic due to the non-radial nature of the magnetic field. Moreover, no standard models exist in the case of magnetic bubbles, analogous to the Leaky dielectric model used for encapsulated electric bubbles \citep{shaw2009electrically,liu2018deformation} to model the magnetoelastic interface. This is primarily because interface magneto-elasticity is not yet sufficiently developed to be applied to bubble dynamics.

To model the behavior of encapsulated magnetic microbubbles, a theory for thin magneto-elastomers is essential. However, there are only a few papers on the magneto-elastic shell theory for such materials. A simpler approach is the membrane theory for magneto-elastomers \citep{barham2007finite,barham2012magnetoelasticity}, which neglects the bending energy of membranes. While this simplification is technically inaccurate, as bubble deformation involves changes in curvature, it can serve as a reasonable approximation due to the low thickness-to-curvature ratio of the bubble, approximately $\mathcal{O}(10^{-3})$. In our study, the encapsulation thickness is taken to be in the range of $10-20~{\rm nm}$. This makes the membrane theory applicable for bubble radii in the range $10-20~\mu{\rm m}$. In this paper, thin membrane theory is adopted to derive mode shape equations by considering nonlinear radial oscillations and linear shape mode oscillations. Using the approximations from \citep{liu2012surface}, the natural frequencies of the shape modes are calculated for the coil field case. A stability diagram is then constructed based on driving pressure amplitude and frequency for both coil and dipole fields, allowing for prediction of the model's range of validity.

The paper is organized as follows. Section \ref{s:2} provides the mathematical formulation of the problem, including the membrane theory for magneto-elastomers, the kinematics of the bubble surface, magnetic field-induced forces, fluid dynamics, and the final governing equations. In Section \ref{s:3}, the natural frequency of the $k^{\rm th}$ mode is derived using the boundary layer approximation for the coil field. In Section \ref{s:4}, a discussion is made on the nature of instability, and a stability criterion for linear non-autonomous systems is presented to validate the computational results. Section \ref{s:5} presents stability diagrams for static current, time-varying current, and dipole strength, illustrating the relationship between driving pressure amplitude and frequency and discussing the corresponding sources of instability. Variations across material parameters and the location of the dipole are examined to estimate the model’s range of validity. Also, the influence of these parameters on the time series of radial and second mode is analyzed. Finally, Section \ref{s:6} offers a summary and conclusion.
\section{Mathematical formulation}\label{s:2}
\subsection {Magneto-elastic equations}
The bubble surface is encapsulated by a lipid/polymer coating infused with magnetic nanoparticles. This facilitates the translation, radial, and non-spherical oscillations of the bubble on the application of a magnetic field. In general, the strain energy density $W$ of magneto elastomer is a function of $\msb{F}$ and $\bs m$  \citep{steigmann2004equilibrium}. Here, $\msb{F}$ is the gradient of the deformed position $\mathbf x$ with respect to reference position $\mathbf  X$, and $\bs m$ is magnetization, which is the magnetic moment per unit volume. For ease of analysis and the range in which the magnetic field is applied $(0.5 {\,\rm T}-1.5 {\,\rm T})$, the interface response is assumed to be weakly magnetic. This simplifies the expression for the free energy of the interface to a sum of elastic energy and magnetic energy as 
\begin{equation}\label{e:2_1}
W=W_e(\msb{F})-\frac{\mu_0\chi}{2}{\bs h}\cdot{\bs h}.
\end{equation}
Here, $\mu_0$ is permeability of free space, $\chi$ is interface susceptibility,  and $\bs{h}$ is induced magnetic field at the interface. This simplifies the  constitutive equations of interface stresses and magnetization as
\begin{equation}\label{e:2_2}
{\bs \sigma}=\frac{\partial W}{\partial \msb F}{\msb F}^{T}-q{\msb I},\quad
{\bs m}=-\frac{1}{\mu_0}\frac{\partial W}{\partial \bs h}=\chi {\bs h}.
\end{equation}
In the above equation, $q$ is constraint pressure associated with incompressibility condition of the membrane, ${\msb F}^{T}$ is transformation of deformation gradient and $\msb I$ is the unit tensor. The induced field can be split as 
\begin{eqnarray}\label{e:2_3}
{\bs h}={\bs h}_a+{\bs h}_s,
\end{eqnarray}
where ${\bs h}_a$ is applied magnetic field and ${\bs h}_s$ is self field which is the secondary field created due to the induced magnetic moment at the interface. Due to the weakly magnetic nature of the membrane, the self field can be ignored. This simplifies the divergence of pondermotive stress $\bs \sigma_{\rm pon}$ which is the sum of magnetic stress
$\bs{\sigma}_{\rm mag}$ and Maxwell stress $\bs{\sigma}_{\rm max}$ as
\begin{align}\label{e:2_4}
\begin{split}
\bs{\sigma}_{\rm max}&=\mu_0(\bs{h\otimes h})-\frac{\mu_0(\bs{h}\cdot \bs{h})}{2} \msb{I},\\
\bs{\sigma}_{\rm mag}&=\mu_0\bs{(h\otimes m)},\\
\bs{\nabla}\cdot\bs{\sigma}_{\rm pon}&=\mu_0\bs{\nabla  h}\cdot \bs{m}\approx\mu_0\chi\bs{\nabla h}_a\cdot\bs{h}_a,
\end{split}
\end{align}
where $\bs\nabla$ is the del operator in the current configuration. The three-dimensional magneto-elastic equation in the absence of applied body forces in the current configuration is  
\begin{align}\label{e:2_5}
\bs{\nabla}\cdot\msb{T}=0,\quad
\msb{T}=\bs{\sigma}+\bs{\sigma}_{\rm pon}.
\end{align}
Using the expression for the divergence of pondermotive stress from~\eqref{e:2_4} in~\eqref{e:2_5} and transforming the interface stress to the reference configuration modifies the above equation as,
\begin{equation}\label{e:2_6}
\bs{\bar{\nabla}}\cdot\textbf{P}+\mu_0\chi\nabla \bs{h}_a\cdot \bs{h}_a=0,
\end{equation}
where $\msb{P}$ is the first Piola stress tensor and $\bs{\bar{\nabla}}$ is del operator in reference configuration. The above equation is also valid at mid surface of the membrane
\begin{equation}\label{e:2_7}
\bs{\bar{\nabla}}\cdot\msb{P}_s+\mu_0\chi\nabla \bs{h}_a\cdot \bs{h}_a=0,
\end{equation}
with $\msb{P}_s$ as the Piola stress evaluated at the membrane mid-surface. Due to the thinness and the very low thickness-to-radius ratio of the bubble surface, approximately $\mathcal{O}(10^{-3})$ the interface can be modeled as a membrane \citep{barham2007finite,barham2012magnetoelasticity}. Let the membrane thickness be $\epsilon$, divergence of the first Piola stress becomes
\begin{equation}\label{e:2_8}
\bs{\bar{\nabla}}\cdot\msb{P}_s=\bs{\bar{\nabla}}_s\cdot\msb{P}_s+\msb{P}_s'\msb{k}.
\end{equation}
Here, $\bs{\bar{\nabla}}_s$ is the two-dimensional surface gradient operator in the reference/undeformed configuration, $\msb{P}_s'$ is the differentiation of first Piola stress in the normal direction, and $\msb{k}$ is normal to the reference surface. Now, expanding first Piola stress in the thickness direction,
\begin{align}\label{e:2_9}
\begin{split}
\msb{P^{+}}&=\msb{P}_{s}+\frac{\epsilon}{2}\msb{P}'_{s},\\
\msb{P^{-}}&=\msb{P}_{s}-\frac{\epsilon}{2}\msb{P}'_{s},\\
\end{split}
\end{align}
where $\msb{P^{\pm}}$ are the limiting value of Piola stress at the top and bottom surface, respectively, from inside, $\msb{k^{\pm}}$ are corresponding exterior normals at top and bottom $\msb{\pm k}$. Adding the traction on top $\bs t_{\text a}^{+}$ and on bottom surface $\bs t_{\text a}^{-}$ the following equations are obtained 
\begin{align}\label{e:2_10}
\begin{split}
\msb{P^{+}k^{+}}&=\alpha\,\bs{t}_a^{+},\\
\msb{P^{-}k^{-}}&=\alpha\,\bs{t}_a^{-},\\
\msb{P^{+}k^{+}+P^{-}k^{-}}&=\epsilon\,\msb{P}'_s\msb{k}.
\end{split}
\end{align}
Here, $\alpha$ is local area dilatation. Substituting the above equations in~\eqref{e:2_7} and rewriting in the current configuration, the following membrane equation is obtained
\begin{align}\label{e:2_11}
\begin{split}
\bs{\nabla}_s\cdot\bs{\sigma}_s+\alpha\frac{\bs{t}_a^+ +\bs{t}_a^-}{\epsilon}+\mu_0\bs{\nabla h}_a\cdot \bs{h}_a=0.
\end{split}
\end{align}
Here, $\bs{\nabla}_s$ is two-dimensional gradient in current/deformed configuration, $\sigma_{s}$ is cauchy stress at mid surface of membrane. In~\eqref{e:2_2} the expression for stress can be represented in terms of principle stretches $\lambda_i, i=1,2,3$ as
\begin{eqnarray}\label{e:2_12}
\bs{\sigma}_s=\sum_{i=1}^{3}\left(\lambda_{i}\frac{\partial W}{\partial\lambda_i}-q\right)\bs{v}_i\otimes \bs{v}_i,
\end{eqnarray}
and the deformation gradient as
\begin{eqnarray}\label{e:2_13}
\msb{F}=\sum_{i=1}^{3}\lambda_i\,\bs{v}_i\otimes \bs{u}_i,
\end{eqnarray}
where $\bs{u}_i$ and $\bs{v}_i$ are the principle stress directions before and after the deformation and $\lambda_i$  is the stretch in the corresponding principle stress direction. Incompressibility assumption of the bubble implies ${\rm det}({\msb F})=1$, which leads to  
$\lambda_3=(\lambda_1\lambda_2)^{-1}$.
A generalized plane stress condition is assumed by which stress vector component $\bs{\sigma}_s\cdot\bs{\hat{n}}$ normal to membrane mid-surface vanishes. This leads the mechanical part of the stress component in the normal direction to zero and the expression for the constraint pressure can be written as $$q=\lambda_3\frac{\partial W}{\partial\lambda_3}.$$
\subsection {Kinematics of bubble surface}
\begin{figure}
    \centering
    \includegraphics[width=1\linewidth]{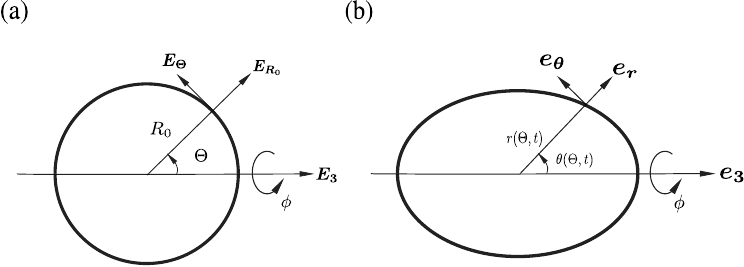}
    \caption{ (a) Deformed and (b) undeformed configurations of a bubble surface.}
    \label{fig:1}
\end{figure}
In this work, only axisymmetric coils with uniform current distribution are considered, which creates axisymmetric magnetic fields, and the applied pressure is radial. Consequently, only the axisymmetric oscillations of the bubble are taken into account.
Considering a spherical bubble of initial radius $R_0$, $(R_0,\Theta)$ in the reference configuration gets transformed to $(r,\theta)$ in the current configuration as shown in figure~\ref{fig:1}. The position of the bubble surface after deformation is expressed in terms of the spherical mode shapes as
\begin{align}\label{e:2_14}
\begin{split}
r(\Theta,t)&=R(t)+\sum_{k=2}^{\infty}a_k(t)P_k(\cos{\Theta}),\\
\theta(\Theta,t)&=\Theta+\sum_{k=1}^{\infty}\frac{b_k(t)}{R(t)}P^1_k(\cos{\Theta}),
\end{split}
\end{align}
with radial mode shape represented by the Legendre polynomial $P_k(\cos{\Theta})$ and  tangential mode shape represented by the associated Legendre polynomial $P^1_k(\cos{\Theta})$. In \eqref{e:2_14} $R(t),a_k(t)$ and $b_k(t)$ are deformed radius, and amplitude of the radial and tangential mode shapes, respectively. The deformed and undeformed mid-surface of encapsulation are,
\begin{align}\label{e:2_15}
\begin{split}
{\mathbf x}&=r(\Theta,t) \cos(\phi)\sin(\theta){\bs{e}}_1+r(\Theta,t) \sin(\phi)\sin(\theta){\bs e}_2+r(\Theta,t) \cos(\theta){\bs e}_3=r(\Theta,t)\bs{e}_r,\\ \\
{\mathbf X}&=R_0 \cos(\phi)\sin(\Theta){\bs E}_1+R_0 \sin(\phi)\sin(\Theta){\bs E}_2+R_0 \cos(\Theta){\bs E}_3=R_0\bs{E}_{R_{0}},
\end{split}
\end{align}
where $\bs E_i$ and $\bs e_i$ are the Cartesian unit basis in the undeformed and deformed interface, respectively. Additionally, $\bs E_{R_0}$ and $\bs e_{r}$ are the unit vector in the radial direction on the undeformed and deformed interface, respectively. The curvilinear basis vectors on the deformed surface are 
\begin{align}\label{e:2_16}
\begin{split}
\bs{g}_1&=\bs{x}_{,\Theta}=r(\Theta,t)_{,\Theta}\,\bs{e_r}+r(\Theta,t)\theta_{,\Theta}\,\bs{e_{\theta}},\\
\bs{g}_2&=\bs{x}_{,\phi}=r(\Theta,t)\sin{\theta}\,\bs{e_{\phi}}.
\end{split}
\end{align}
The undeformed basis vectors are given by
\begin{align}\label{e:2_17}
\begin{split}
\bs{G}_1&=\bs{X}_{,\Theta}=R_{0}\bs{E}_{\Theta}, \quad\quad\quad \bs{G}^1=\frac{\bs{E}_{\Theta}}{R_0},\\
\bs{G}_2&=\bs{X}_{,\phi}=R_0\sin{\phi}\bs{E}_{\phi}, \quad \bs{G}^2=\frac{\bs{E}_{\phi}}{R_0\sin{\phi}},
\end{split}
\end{align}
where the lower and upper subscripts correspond to co-variant and contra-variant vectors. The gradient on the deformed mid-surface can be written as
\begin{equation}\label{e:18}
\bs{\bar{\nabla}x}=\bs{g}_{\alpha}\otimes\bs{G}^{\alpha},
\end{equation}
\begin{equation}\label{e:2_19}
\begin{split}
\bs{\bar{\nabla}x}&=\frac{r(\Theta,t)_{,\Theta}}{R_0}(\bs{e_r}\otimes \bs{E_\theta})+\frac{r(\Theta,t)\theta_{,\Theta}}{R_0}(\bs{e_\theta} \otimes \bs{E_\Theta}) +\frac{r(\Theta,t)\sin{\theta}}{R_0\sin{\Theta}}(\bs{e_{\phi}} \otimes\bs{E_{\phi}}),\\
&=\frac{\sqrt{(r(\Theta,t)_{,\Theta})^2+(r(\Theta,t)\theta_{,\Theta})^2}}{R_0}(\bs{\hat{t}}\otimes\bs{E_{\theta}})+\frac{r(\Theta,t)\sin{\theta}}{R_0\sin{\Theta}}(\bs{e_{\phi}} \otimes\bs{E_{\phi}}),\\
\msb{F}&=\bs{\bar{\nabla}x}+\frac{1}{\lambda_1\lambda_2}(\hat{\bs n}\otimes \bs{e}_r).
\end{split}
\end{equation}
Comparing above equations with~\eqref{e:2_13}, the principle stretches in meridional and circumferential directions become
\begin{align}\label{e:2_20}
\begin{split}
\lambda_1&=\frac{\sqrt{(r(\Theta,t)_{,\Theta})^2+(r(\Theta,t)\theta_{,\Theta})^2}}{R_0}\approx\frac{r\theta^{'}}{R_0}\approx\frac{R(t)}{R_0}+\frac{a_k(t)}{R_0}P_k+\frac{b_k(t)}{R_0}\frac{\partial P^1_k}{\partial\Theta},\\
\lambda_2&=\frac{r(\Theta,t)\sin(\theta)}{R_0\sin(\Theta)}\approx\frac{R(t)}{R_0}+\frac{a_k(t)}{R_0}P_k+\frac{b_k(t)}{R_0}\cot(\Theta)P^1_k.
\end{split}
\end{align}
Similarly, the normal $(\bs{\hat{n}})$ and the tangent $(\bs{\hat{t}})$ vectors on the deformed surface are
\begin{align}\label{e:2_21}
\begin{split}
\bs{\hat{n}}\approx \bs{e}_{r}-\frac{a_k(t)}{R(t)}P^1_k \bs{e}_{\theta},\\
\bs{\hat{t}}\approx \bs{e}_{\theta}+\frac{a_k(t)}{R(t)}P^1_k \bs{e}_{r}.
\end{split}
\end{align}
From the expression for normal, the principle curvature of the deformed surface is obtained as
\begin{align}\label{e:2_22}
\bs{\kappa}&=\bs{\nabla\hat{n}}=\kappa_1\,\bs{e}_{\theta}\otimes \bs{e}_{\theta}+\kappa_2\,\bs{e}_{\phi}\otimes \bs{e}_{\phi},
\end{align}
where the principal curvature components $\kappa_1$ and $\kappa_2$ are given by
\begin{align}\label{e:2_23}
\begin{split}
\kappa_1&\approx\frac{1}{R(t)}-\sum_{k=2}^{\infty}\frac{a_k(t)}{R(t)^2}\left(P_k+\frac{{\rm d}^2P_k}{{\rm d}\theta^2}\right),\\
\kappa_2&\approx\frac{1}{R(t)}-\sum_{k=2}^{\infty}\frac{a_k(t)}{R(t)^2}\left(P_k+\frac{\cos(\theta)}{\sin(\theta)}P^1_k\right).
\end{split}
\end{align}
The membrane is assumed to follow Mooney-Rivlin constitutive law as follows
\begin{equation}\label{e:2_24}
W_e=C_1\left(\lambda_1^2+\lambda_2^2+\frac{1}{\lambda_1^2\lambda_2^2} -3\right) +C_2\left(\frac{1}{\lambda_1^2}+\frac{1}{\lambda_2^2}+\lambda_1^2\lambda_2^2-3\right).
\end{equation}
The in-plane mechanical stress becomes
\begin{align}\label{e:2_25}
\begin{split}
\sigma_{1}&=\lambda_1\frac{\partial W_e}{\partial\lambda_1}=2C_1\bigg(\lambda_1^2-\frac{1}{\lambda_1^2\lambda_2^2}\bigg)+2C_2\bigg(\lambda_1^2\lambda_2^2-\frac{1}{\lambda_1^2}\bigg),\\
\sigma_{2}&=\lambda_2\frac{\partial W_e}{\partial\lambda_2}=2C_1\bigg(\lambda_2^2-\frac{1}{\lambda_2^2\lambda_1^2}\bigg)+2C_2\bigg(\lambda_2^2\lambda_1^2-\frac{1}{\lambda_2^2}\bigg).
\end{split}
\end{align}
The divergence of in-plane stress becomes \citep{pozrikidis2001effect},
\begin{align}\label{e:2_26}
\begin{split}
(\bs{\nabla}\cdot\bs{\sigma})_n&=-\kappa_1\sigma_1-\kappa_2\sigma_2,\\
(\bs{\nabla}\cdot\bs{\sigma})_t&=\frac{\partial\sigma_1}{\partial s}+\frac{1}{d}\frac{\partial d}{\partial s}(\sigma_1-\sigma_2),
\end{split}
\end{align}
where $d$ is the horizontal distance of point in the membrane from the axis of revolution of bubble.
\begin{equation}\label{e:2_27}
d=R(t)\sin(\theta)+a_k(t)\sin(\theta)P_k+b_k(t)\cos(\theta)P^1_k.
\end{equation}
Substituting the stretches and $x$ in the divergence expression, we get
\begin{align}\label{e:2_28}
\begin{split}
F^{\rm mem}_n&=\frac{\left(\bs{\nabla}\cdot{\bs \sigma}_s\right)_n}{\lambda_1\lambda_2} \approx\mathcal{S}_0+\mathcal{S}_1a_k(t)+\mathcal{S}_2b_k(t),\\
F^{\rm mem}_t&=\frac{\left(\bs{\nabla}\cdot{\bs \sigma}_s\right)_t}{\lambda_1\lambda_2}\approx\mathcal{S}^1_1a_k(t)+\mathcal{S}^1_2b_k(t).
\end{split}
\end{align}

\begin{align}\label{e:2_29}
\begin{split}
\mathcal{S}_0&=C_1\left(\frac{-4R_0^6}{R^{7}}+\frac{4}{R}\right)+C_2\left(-\frac{4R_0^{4}}{R^{5}}+\frac{4R}{R_0^2}\right),\\
\mathcal{S}_1&=C_1\left(-(2k^2+2k-28)\frac{R_0^6}{R^8}+\frac{2k^2+2k-4}{R^2}\right)\\&\quad+C_2\left(-(2k^2+2k-20)\frac{R_0^4}{R^6}+\frac{2k^2+2k+4}{R_0^2}\right),\\
\mathcal{S}_2&=C_1\left(-12(k^2+k)\frac{R_0^6}{R^8}\right)+C_2\left(-(8k^2+8k)\frac{R_0^4}{R^6}-\frac{4k^2+4k}{R_0^2}\right),\\
\mathcal{S}^{1}_{1}&=C_1\left(-\frac{8R_0^6}{R^8}-\frac{4}{R^2}\right)+C_2\left(-\frac{4R_0^4}{R^6}-\frac{8}{R_0^2}\right),\\
\mathcal{S}^{1}_{2}&=C_1\left((4k^2+4k)\frac{R_0^6}{R^8}+\frac{4k^2+4k-4}{R^2}\right)+C_2\left((4k^2+4k-4)\frac{R_0^4}{R^6}+\frac{4k^2+4k}{R_0^2}\right).
\end{split}
\end{align}
\subsection{Magnetic field and forces}
\begin{figure}
    \centering
    \includegraphics[width=1\linewidth]{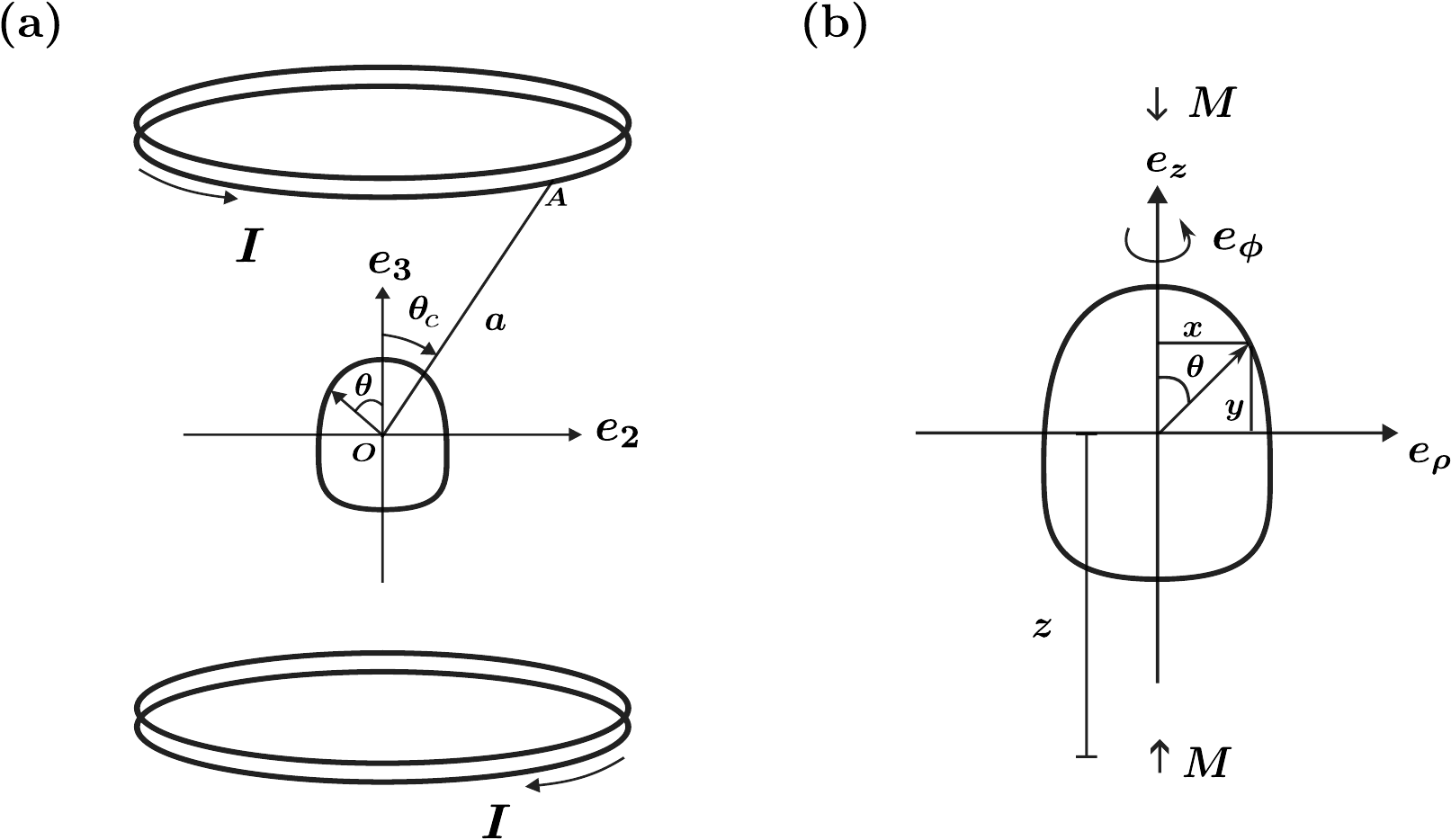}
    \caption{Schematics for magnetic field (a) Two coils placed symmetrically about ${\bs e}_1-{\bs e}_2$ plane carrying currents in opposite direction, (b) Two dipoles placed symmetrically about $\bs{e}_\rho-\bs{e}_\phi$ plane.}
    \label{fig:2}
\end{figure}
\subsubsection{Coil field}
Two coils are symmetrically placed above and below a bubble, each carrying current $I$ in opposite directions and having $N_1$ turns as shown in figure~\ref{fig:2}a. While the Biot-Savart law offers a straightforward approach to calculate the resulting magnetic field, it often involves the use of complex elliptic integrals, making the computation cumbersome. To simplify the analysis, an alternative method proposed by \citep{boridy1989magnetic} is used to calculate magnetic fields for axially symmetric systems. The generalized expression for axisymmetric magnetic fields is
\begin{align}\label{e:2_30}
\begin{split}
\bs{h}_a&=h_{ar}\bs{e}_r+h_{a\theta}\bs{e}_{\theta},\\
h_{ar}&=\sum_{n=1}^{\infty}H_n\left(\frac{r}{a}\right)^{n-1}P_{n}(\cos{\theta}),\\
h_{a\theta}&=\sum_{n=1}^{\infty}\frac{H_n}{n}\left(\frac{r}{a}\right)^{n-1}P_{n}^{1}(\cos{\theta}),
\end{split}
\end{align}
where $r$ and $\theta$ are radial and angular position of any point in the $r-z$ plane, $a$ is the distance from the bubble $O$ to any point on the coil circumference $A$  and $\theta_c$ is the angle between axis of symmetry $\bs{e}_3$ and the vector along $OA$. For the case of coils carrying currents in opposite directions placed symmetrically above and below the equator, the odd terms vanish in~\eqref{e:2_30} and only even terms are sustained. The expressions for even coefficients are given below,
\begin{equation}\label{e:2_31}
H_{2n}=-\frac{I}{a}N_{1}\sin{\theta_c}\,P^{1}_{2n}(\cos{\theta_c}).
\end{equation}
By choosing the value of $\theta_c$ as a positive root of $P^{1}_4$, the co-efficient $H_4$ can be made zero. As we are interested in fields and field gradients in the region of the bubble, the $r/a$ ratio is very small. Hence, the subsequent terms $H_{6}, H_{8},...$ in~\eqref{e:2_30} can be ignored and only leading terms can be considered, resulting in simplified expressions for magnetic fields, as
\begin{align}\label{e:2_32}
\begin{split}
h_{ar}&=H_{2}\frac{r}{2a}(3\cos^2{\theta}-1),\\
h_{a\theta}&=H_{2}\frac{r}{2a}(-3\sin{\theta}\cos{\theta}).
\end{split}
\end{align}
In-spherical co-ordinates the magnetic forces become,
\begin{align}\label{e:2_33}
\begin{split}
F^{\rm mag}_{r}&=(\mu_0\chi\bs\nabla\bs{h}_{a}\cdot\bs{h}_a)_r=\mu_0\chi\left(h_{ar}\frac{\partial h_{ar}}{\partial r}+\frac{h_{a\theta}}{r}\frac{\partial h_{ar}}{\partial\theta}-\frac{h_{a\theta}^2}{r}\right),\\&=\mu_0\chi\frac{H_{2}^{2}r}{4a^2}(3\cos^2{\theta}+1),\\
F^{\rm mag}_{\theta}&=(\mu_0\chi\bs\nabla\bs{h}_{a}\cdot\bs{h}_a)_\theta=\mu_0\chi\left(h_{ar}\frac{\partial h_{a\theta}}{\partial r}+\frac{h_{a\theta}}{r}\frac{\partial h_{a\theta}}{\partial\theta}+\frac{h_{ar}h_{a\theta}}{r}\right),\\&=\mu_0\chi\frac{H_{2}^{2}r}{4a^2}(-3\sin{\theta}\cos{\theta}).
\end{split}
\end{align}
\subsubsection{Dipole field}
The expression for magnetic field generated due to dipole is~\citep{latulippe2014dipole},
\begin{align}\label{e:2_34}
\bs{h}_a=M\left[\frac{3(\bs{r}\cdot{\bs{e}}_z)\bs{r}}{\norm{\bs{r}}^5}-\frac{{\bs{e}}_z}{\norm{\bs{r}}^3}\right].
\end{align}
Here, $M$ is the magnetic dipole strength, $\bs{r}$ is the vector position of the bubble surface from the dipole, and $\norm{\cdot}$ denotes Euclidean norm. Assuming that the dipole is placed sufficiently far away from the bubble $z\gg x$ and $z\gg y$, the above expression for the magnetic field gets simplified as,
\begin{align}\label{e:2_35}
\bs{h}_a=\frac{3Mx}{z^4}\bs{e}_\rho+\frac{2M}{z^3}\bs{e}_z.
\end{align}
As two dipoles are place symmetrically about $\bs{e}_r$ the field components along $\bs{e}_z$ cancels and component along $e_r$ adds up resuluting in,
\begin{align}\label{e:2_36}
\bs{h}_a=\frac{6Mx}{z^4}\bs{e}_\rho.
\end{align}
The forcing on the bubble surface due to dipole field is,
\begin{align}
\mu_0\chi_s\bs{\nabla h}_a\cdot\bs{h}_a&=\frac{36\mu_0\chi_sM^2x}{z^8}\bs{e}_\rho\label{e:2_37},\\
\bs{F}^{mag}&=\left(\frac{1}{\lambda_1\lambda_2}\right)\frac{36\mu_0\chi M^2}{z^8}r\sin{\theta}\bs{e}_\rho\label{e:2_38}.
\end{align}
The calculations for the dipole field are made using cylindrical co-ordinates as shown in figure~\ref{fig:2}b. Substituting for $r$ and $\theta$ the deformed radial and meridional position from \eqref{e:2_14} in the above equations, the radial and meridional components of magnetic force are obtained. Once these components are obtained in spherical coordinates, the dot product of the magnetic forces with the normal and tangential vectors is calculated. This gives us the corresponding force components in the normal and tangential directions as
\begin{align}\label{e:2_39}
\begin{split}
F^{\rm mag}_{n}&=\frac{1}{\lambda_1\lambda_2}\left(F^{\rm mag}_{r}-\frac{a_k(t)P^1_k}{R(t)}F^{\rm mag}_{\theta}\right),\\
F^{\rm mag}_{t}&=\frac{1}{\lambda_1\lambda_2}\left(F^{\rm mag}_{\theta}+\frac{a_k(t)P^1_k}{R(t)}F^{\rm mag}_{r}\right).
\end{split}
\end{align}
The above equations are linearized in $a_k$ and $b_k$ to be consistent with the present formulation. The final equations after using the recurrence relations of Legendre polynomials are 
\begin{align}\label{e:2_40}
\begin{split}
F^{\rm mag}_{n}&=\mathcal{H}_0+f_1b_1P_1+\mathcal{H}_1P_2+\mathcal{H}\left[\sum_{k=2}^{\infty}\mathcal{F}_{1}(k)a_kP_k +\sum_{k=2}^{\infty}\mathcal{F}_{2}(k)b_kP_k\right.\\&+\left.\sum_{k=0}^{\infty}\mathcal{F}_{3}(k)a_{k+2}P_k +\sum_{k=0}^{\infty}\mathcal{F}_{4}(k)b_{k+2}P_k+\sum_{k=4}^{\infty}\mathcal{F}_{5}(k)a_{k-2}P_k+\sum_{k=4}^{\infty}\mathcal{F}_{6}(k)b_{k-2}P_k\right],\\
F^{\rm mag}_{t}&=f^1_1b_1P^1_1+\mathcal{H}^1_2P^1_2+\mathcal{H}\left[\sum_{k=2}^{\infty}\mathcal{F}^{1}_{1}(k)a_kP^1_k +\sum_{k=2}^{\infty}\mathcal{F}^{1}_{2}(k)b_kP^1_k\right.\\&+\left.\sum_{k=0}^{\infty}\mathcal{F}^{1}_{3}(k)a_{k+2}P^1_k+\sum_{k=0}^{\infty}\mathcal{F}^{1}_{4}(k)b_{k+2}P^1_k+\sum_{k=4}^{\infty}\mathcal{F}^{1}_{5}(k)a_{k-2}P^1_k+\sum_{k=4}^{\infty}\mathcal{F}^{1}_{6}(k)b_{k-2}P^1_k\right],\\
\end{split}
\end{align}
where the expressions for $\mathcal{H}$, $\mathcal{H}_i$, $\mathcal{H}^1_i$,$\mathcal{F}_i$ and $\mathcal{F}^1_i$ are given in Appendix~\ref{A:1} for both coil and dipole field. It is evident from \citet{liu2012surface,liu2018deformation} that applied acoustic pressure primarily excites the radial mode. If we disregard any initial disturbances to the non-spherical modes, the applied magnetic field excites the radial mode as well as the $a_2$ and $b_2$ modes. Additionally, the forcing on the $k^{\rm th}$ mode depends on the neighboring modes, specifically the $k-2$ and $k+2$ modes. As a result, under the influence of the magnetic field generated by both the coil and dipole, only the even-numbered modes are triggered. This allows us to ignore the translations and odd-numbered modes in the equations for mode shapes.
\subsection{Fluid equations}
The assumption is made that the flow field inside the bubble is negligible and fluid flow is considered only outside the bubble, governed by the incompressible Navier–Stokes equations
\begin{align}\label{e:2_41}
\begin{split}
{\bs \nabla}\cdot{\bs u}&=0,\\
\rho\frac{\partial \bs{u}}{\partial t}+\rho[{\bs u}\cdot{\bs \nabla}]\bs{u}&=-\bs{\nabla} p+\eta\bs{\nabla}\cdot\bs{(\nabla u+\nabla u^T)},
\end{split}
\end{align}
where $\bs{u}$ and $p$ are fluid velocity and pressure, respectively. Here, $\rho$ and $\eta$ represent the density and viscosity of the fluid, respectively. The non-spherical oscillations are assumed to be small compared to radial oscillations, and the velocity is split into potential and viscous flow \citep{prosperetti1977viscous} as follows:
\begin{equation}\label{e:2_42}
\bs{u}=\bs{u}_p+\bs{u}_v.
\end{equation}
The velocity of potential flow is calculated by substituting $\eta=0$. The generalized solution for the axisymmetric Laplace equation for incompressible flow is
\begin{equation}\label{e:2_43}
\psi=\frac{C_0P_0}{r}+\sum_{k=2}^{\infty}\frac{C_kP_k}{r^{k+1}},
\end{equation}
\begin{equation}\label{e:2_44}
\bs{u}_p=\frac{\partial\psi}{\partial r}\bs{e}_r+\frac{1}{r}\frac{\partial\psi}{\partial\theta}\bs{e}_\theta.
\end{equation}
The fluid velocity continuity at the interface in the normal direction is given by
\begin{align}\label{e:2_45}
\begin{split}
S&=r-R-a_kP_k,\\
0&=\frac{\partial S}{\partial t}+(\bs{\nabla\psi})\cdot\bs{\nabla} S.
\end{split}
\end{align}
Expanding the terms in \eqref{e:2_45} gives
\begin{align}\label{e:2_46}
\begin{split}
-\dot{R}+\sum_{k=2}^{\infty} -\dot{a}_{k}P_k-\frac{C_0}{R^2}+\frac{2C_0a_kP_k}{R^3}-\frac{(k+1)C_kP_k}{R^{k+2}}=0.
\end{split}
\end{align}
The coefficients in \eqref{e:2_46} are determined by using the orthogonality property of Legendre polynomials. These conditions are given by collecting\\\\
\noindent $0^{\rm th}$ order coefficient:
\begin{equation}\label{e:2_47}
\begin{split}
\dot{R}+\frac{C_0}{R^2}=0,\\
C_0=-R^2\dot{R},
\end{split}
\end{equation}
\newline
$k^{\rm th}$ order coefficient:
\begin{equation}\label{e:2_48}
\begin{split}
-\dot{a}_k&+\frac{2C_0a_k}{R^3}-\frac{C_k(k+1)}{R^{k+2}}=0,\\
C_k&=-\frac{R^{k+2}}{k+1}\left(\dot{a}_k+\frac{2a_k\dot{R}}{R}\right).
\end{split}
\end{equation}

In the Navier-Stokes equation viscosity is substituted as zero and integrated by substituting potential flow velocity and gives pressure developed due to potential flow as
\begin{equation}\label{e:2_49}
p_p=p_\infty-\rho\left(\frac{\partial\psi}{\partial t}+\frac{1}{2}|\bs{\nabla}\psi|^2\right),
\end{equation}
The viscous correction to the potential flow is obtained by solving the linearized Navier-Stokes equation in velocity-vorticity form
\begin{align}\label{e:2_50}
\begin{split}
\frac{\partial\boldsymbol{\omega}}{\partial t}+\bs{(\nabla\times\omega\times u)}&=-\nu\bs{\nabla\times(\nabla\times\omega)},\\
\bs{\omega}&=\bs{\nabla\times u}.
\end{split}
\end{align}
Due to axisymmetric motion of bubble only toroidal component of vortex is considered as
\begin{align}\label{e:2_51}
\bs{\omega}=\bs{\nabla\times}(T_kP_k\bs{e}_r).
\end{align}
Here, $T_k$ is toroidal component of vorticity~\cite{prosperetti1977viscous}, substituting above in Navier-stokes equation gives~\citep{liu2012surface,liu2018deformation}.
\begin{align}\label{e:2_52}
\rho\frac{\partial T_k}{\partial t}+\rho\frac{\partial}{\partial r}\left(\frac{\dot{R}R^2T_k}{r^2}\right)-\eta\frac{\partial^2T_k}{\partial r^2}+\eta k(k+1)r^{-2}T_k=0.
\end{align}
 The viscous velocity and viscous pressure are obtained as
\begin{align}
\bs{u}_v&=\left(\sum_{k=2}^{\infty}T_kP_k-\frac{\partial\Phi}{\partial r}\right){\bs e_r}-\frac{1}{r}\frac{\partial\Phi}{\partial\theta}\bs{e}_\theta\label{e:2_53},\\
p_v&=\sum_{k=1}^{\infty}k\left[\mu\frac{T_k}{R}+\frac{\rho\dot{R}}{R}\int_{R}^{\infty}\left[\frac{R^3}{s^3}-1\right]\frac{R^k}{s^k}T_k \,{\rm d}s\right]P_k\label{e:2_54},\\
\Phi&=\sum_{k=2}^{\infty}P_k\bigg[\bigg(-\frac{k+1}{2k+1}\int_{R}^{\infty}s^{-k}T_k \,{\rm d}s+\frac{k+1}{2k+1}\int_{R}^{r}s^{-k}T_k \,{\rm d}s\bigg)r^{k}\nonumber\\&
\;-\bigg(\frac{k}{2k+1}R^{2k+1}\int_{R}^{\infty}s^{-k}T_k\,{\rm d}s -\frac{k}{2k+1}\int_{R}^{r}s^{k+1}T_k\,{\rm d}s\bigg)r^{-(k+1)}\bigg]\label{e:2_55}.
\end{align}
The viscous stress due to this flow on the bubble surface becomes,
\begin{equation}\label{e:2_56}
\msb{T^{\rm fluid}_{\rm ex}}=\eta(\bs{\nabla u+\nabla u^T}).
\end{equation}
The components of this flow in normal and tangential directions becomes
\begin{align}
\bs{\hat{n}}\cdot\msb{T}^{\rm fluid}_{\rm ex}\cdot\bs{\hat{n}}&=\frac{-4\eta\dot{R}}{R}+\sum_{k=2}^{\infty}2\mu\bigg[(k+2)\frac {\dot{a}_k}{R}-2(k-1)\frac{\dot{R}}{R^2}a_k\bigg.\nonumber\\&\hspace{1.7in}\bigg.+k(k+1)R^{k-2}\int_{R}^{\infty}s^{-k}T_k\,{\rm d}s\bigg]P_k\label{e:2_57},\\
\bs{\hat{t}}\cdot\msb{T}^{\rm fluid}_{\rm ex}\cdot\bs{\hat{n}} &=\sum_{k=2}^{\infty}2\eta\bigg[\frac{k+2}{k+1}\frac{\dot{a}_k}{R}+\frac{1-k}{k+1}\frac{\dot{R}a_k}{R^2}-\frac{T_k}{2R}-R^{k-2}\int_{R}^{\infty}s^{-k}T_k \,{\rm d}s\bigg]P^1_k\label{e:2_58}.
\end{align}
\subsection{{Final Equations}}
The traction on the lower and upper surfaces of the membrane are
\begin{align}\label{e:2_59}
\begin{split}
\bs{t}_a^+&=-p_{\rm ex}\bs{\hat{n}}+\msb{T}^{\rm fluid}_{\rm ex}\bs{\hat{n}},\\
\bs{t}_a^-&=\,\,\,p_{\rm in}\bs{\hat{n}}.
\end{split}
\end{align}
Dividing ~\eqref{e:2_11} by $-\alpha$ and substituting $F^{\rm mem}$, $F^{\rm mag}$, $\bs{t}^{+}_a$ and $\bs{t}^{-}_{a}$ in the governing equation stress balance at the interface, following equations are obtained:
\begin{equation}\label{e:2_60}
\begin{split}
F^{\rm mem}_{n}+\frac{p_{\rm ex}}{\epsilon}-\frac{p_{\rm in}}{\epsilon}-\frac{\bs{\hat{n}}\cdot\msb{T^{\rm fluid}_{\rm ex}}\bs{\hat{n}}}{\epsilon}-F^{\rm mag}_{\rm rad}=0,\\
F^{\rm mem}_{t}-\frac{\bs{\hat{t}}\cdot\msb{T^{\rm fluid}_{\rm ex}}\bs{\hat{n}}}{\epsilon}-F^{\rm mag}_{\rm tan}=0.
\end{split}
\end{equation}
From the expression for the divergence of stresses at the interface, tractions from above and below it can be observed that they are well separated in terms of Legendre or associated Legendre polynomials. The orthogonality property of these polynomials can be utilized to decompose the equation into separate modes.
\begin{equation}
\left. \begin{array}{ll}
\displaystyle\int_{-1}^{1}P_{n}^{k}P_m^kdx=0
  \quad \mbox{when\ }\quad [m\not= n],\\[8pt]
\displaystyle\int_{-1}^{1}P_{n}^{k}P_m^kdx=\frac{2}{2n+1}\frac{(n+m)!}{(n-m)!} 
  \quad \mbox{when\ }\quad [m=n],
 \end{array}\right\}
  \label{symbc}
\end{equation}
Multiplying \eqref{e:2_60}$_1$ by $P_k$ and integrating the equations for radial, and $k^{\rm th}$ mode shape yields the following equation
\newline
\begin{equation}\label{e:2_62}
\rho\left(R\ddot{R}+\frac{3}{2}\dot{R}^2\right)+p_\infty-p_{in}+4\eta\frac{\dot{R}}{R}+F^{\rm mem}_{n0}\mathbf{-\mathcal{H}_0-\mathcal{H}[\mathcal{F}_3(0)a_2(t)+\mathcal{F}_4(0)b_2(t)]=0},
\end{equation}
\newline
\begin{align}\label{e:2_63}
&\rho\left(\frac{R\ddot{a}_2}{3}+\dot{R}\dot{a}_2-\frac{\ddot{R}}{3}a_2+\frac{2\dot{R}}{R}\int_{R}^{\infty}\left[\frac{R^3}{s^3}-1\right]\frac{R^2}{s^2}T_2 \,{\rm d}s\right) +\frac{2\eta T_2}{R}\nonumber\\&\qquad+ {2\eta}\left[\frac{4\dot{a}_2}{R}+\frac{2\dot{R}a_2}{R^2}-6\int_{R}^{\infty}s^{-2}T_2 \,{\rm d}s\right]+F^{\rm mem}_{n2}\mathbf{-\mathcal{H}}_2\\&\qquad\mathbf{-\mathcal{H}[\mathcal{F}_1(2)a_2(t)+\mathcal{F}_2(2)b_2(t) +\mathcal{F}_3(2)a_4(t)+\mathcal{F}_4(2)b_4(t)]}=0\nonumber,
\end{align}
\begin{align}\label{e:2_64}
\begin{split}
&\rho\left(\frac{R\ddot{a}_k}{k+1}+\left[\frac{3\dot{R}}{k+1}+\frac{2(k+2)\eta}{\rho R}\right]\dot{a}_k+\left[\frac{-(k-1)\ddot{R}}{k+1}+\frac{4(k-1)\eta}{\rho}\frac{\dot{R}}{R^2}\right]a_k
+\frac{k\eta T_k}{\rho R}\right.\\&\left.+k\frac{\dot{R}}{R}\int_{R}^{\infty}\left[\frac{R^3}{s^3}-1\right]\frac{R^k}{s^k}T_k \,{\rm d}s -\frac{2\eta}{\rho}k(k+1)\int_{R}^{\infty}s^{-k}T_k \,{\rm d}s\right)
+F^{\rm mem}_{nk}\mathbf{-\mathcal{H}\left[\mathcal{F}_{1}(k)a_k(t)\right.}\\&\mathbf{\left.+\mathcal{F}_{2}(k)b_k(t)+\mathcal{F}_{3}(k)a_{k+2}(t)+\mathcal{F}_{4}(k)b_{k+2}(t)+\mathcal{F}_{5}(k)a_{k-2}(t)+\mathcal{F}_{6}(k) b_{k-2}(t)\right]}=0.
\end{split}
\end{align}
Similarly, orthogonality conditions are used to simplify \eqref{e:2_60}$_2$ to obtain the equations for $b_k$ as 
\begin{align}\label{e:2_65}
&2\eta\left[\frac{k+2}{k+1}\frac{\dot{a}_k}{R}+\frac{1-k}{k+1}\frac{\dot{R}a_k}{R^2}-\frac{T_k}{2R}-R^{k-2}\int_{R}^{\infty}s^{-k}T_k \,{\rm d}s\right]+F^{\rm mem}_{tk}\mathbf{-\mathcal{H}^1_2\delta_{k2}-\mathcal{H}[\mathcal{F}^{1}_{1}(k)a_k(t)\nonumber}\\&\mathbf{+\mathcal{F}^{1}_{2}(k)b_k(t)+\mathcal{F}^{1}_{3}(k)a_{k+2}(t) +\mathcal{F}^{1}_{4}(k)b_{k+2}(t)+\mathcal{F}^{1}_{5}(k)a_{k-2}(t)+\mathcal{F}^{1}_{6}(k) b_{k-2}(t)]}=0.
\end{align}
\newline
Along with this no-slip boundary condition at the surface of the bubble in tangential direction is used
\begin{equation}\label{e:2_66}
u_{\theta}=r\frac{\partial\theta}{\partial t}.
\end{equation}
Substituting the deformed radial and angular positions of bubble surface from \eqref{e:2_14} in above equation it gets simplified as
\begin{equation}\label{e:2_67}
-\frac{\dot{a}_k}{k+1}-\frac{2\dot{R}a_k}{(k+1)R}+R^{k-1}\int_{R}^{\infty}s^{-k}T_k \,{\rm d}s=\dot{b}_k-\frac{\dot{R}b_k}{R}.
\end{equation} 
It can be observed from equations~\eqref{e:2_62},~\eqref{e:2_63} and \eqref{e:2_64} that the radial oscillations are directly influenced by both the applied pressure and the magnetic field. In contrast, the $a_2$ shape mode is directly driven by the magnetic field, while its response to pressure is indirect, arising from the coupling between the radial and shape mode equations. All other even shape modes are similarly excited through the coupling between equations, rather than by direct external forcing.
\section{Natural frequency calculation for coil field}\label{s:3}
To estimate the natural frequency of magnetic bubbles the thin boundary layer approximation for acoustic streaming near a surface is used similar to the approach of
\citet{liu2012surface}. This simplifies terms in~\eqref{e:2_67} as
\begin{align}\label{e:3_1}
\begin{split}
&\delta_k=\text{min}\left(\sqrt{\frac{\eta}{\rho\omega}},\frac{R}{2k}\right),\\
&\int_{R}^{\infty}\left[\frac{R^3}{s^3}-1\right]\bigg(\frac{R}{s}\bigg)^kT_k(s,t){\,\rm d}s\approx0,\\
&\int_{R}^{\infty}s^{-k}T_k(s,t){\,\rm d}s\approx R^{-k}T(R,t)\delta_k.
\end{split}
\end{align}
To calculate the natural frequency of $k^{\rm th}$ mode, in~\eqref{e:2_64} and~\eqref{e:2_65} above approximations are substituted.
\begin{align}\label{e:3_2}
\begin{split}
\ddot{a}_k+C_{a1}\dot{a}_k+C_{b1}\dot{b}_k+C_aa_k+C_bb_k+\Psi_1(a_{k+2},b_{k+2},a_{k-2},b_{k-2})=0,\\
D_{b1}\dot{b}_k+D_{a1}\dot{a}_k+D_aa_k+D_bb_k+\Psi_2(a_{k+2},b_{k+2},a_{k-2},b_{k-2})=0,
\end{split}
\end{align}
where coefficients are,
\begin{align}\label{e:3_3}
\begin{split}
C_a&=\frac{\mathcal{S}_1}{\rho R}-\frac{H}{4\rho R^3}\mathcal{F}_1-(k-1)\frac{\ddot{R}}{R}+\left[(k-1)(k+1)(k+2)+\frac{k(k+2)R}{\delta_k}\right]\frac{2\eta\dot{R}}{\rho R^3},\\
C_b&=\frac{\mathcal{S}_2}{\rho R}-\frac{H}{4\rho R^3}\mathcal{F}_2,\\
D_a&=-\frac{\mathcal{S}^1_1}{\rho R}+\frac{H}{4\rho R^3}\mathcal{F}^1_1-(k+1+\frac{R}{\delta_k})\frac{2\eta\dot{R}}{R^2},\\
D_b&=-\frac{\mathcal{S}^1_2}{\rho R}+\frac{H}{4\rho R^3}\mathcal{F}^1_2+(k+1)(2+\frac{R}{\delta_k})\frac{\eta\dot{R}}{R^2},
\end{split}
\end{align}
where $H=\frac{\mu_0\chi_SH_2^2R_0^2}{a^2}$. The other coefficients can be found in \citep{liu2012surface}. Based on the arguments presented in the same paper, the magnitudes of $D_{a1}$ and $D_{b1}$ are negligible compared to other terms in the second equation of~\eqref{e:3_2}. To calculate the natural frequency for the $k^{\rm th}$ mode, we can assume $D_{a1}\approx 0$ and $D_{b1}\approx 0$. The second equation of~\eqref{e:3_2} is linearized in $R$ and written in matrix form as
\begin{equation}\label{e:3_4}
A_a
\begin{Bmatrix}
 a_k   
\end{Bmatrix}
+B_b
\begin{Bmatrix}
 b_k   
\end{Bmatrix}+
\begin{Bmatrix}
-\frac{H}{4R_0}\\
\vdots\\
\end{Bmatrix}=0.
\end{equation}
\begin{align}\label{e:3_5}
\begin{split}
A_{a}=
\left[\begin{smallmatrix}
\cdots&\cdots&\cdots&\cdots&\cdots&\cdots&\cdots&\cdots\\
\vdots&-\frac{H\mathcal{F}^1_5(k)}{4R_0^2}&0&\mathcal{S}^1_1(k)-\frac{H\mathcal{F}^1_1(k)}{4R_0^2}&0&-\frac{H\mathcal{F}^1_3(k)}{4R_0^2}&0&0&\vdots\\
\vdots&0&-\frac{H\mathcal{F}^1_5(k+1)}{4R_0^2}&0&\mathcal{S}^1_1(k+1)-\frac{H\mathcal{F}^1_1(k+1)}{4R_0^2}&0&-\frac{H\mathcal{F}^1_3(k+1)}{4R_0^2}&0&\vdots\\
\cdots&\cdots&\cdots&\cdots&\cdots&\cdots&\cdots&\cdots\\
\end{smallmatrix}\right]
\end{split}
\end{align}
\begin{align}\label{e:3_6}
\begin{split}
B_{b}=
\left[\begin{smallmatrix}
\cdots&\cdots&\cdots&\cdots&\cdots&\cdots&\cdots&\cdots\\
\vdots&-\frac{H\mathcal{F}^1_6(k)}{4R_0^2}&0&\mathcal{S}^1_2(k)-\frac{H\mathcal{F}^1_2(k)}{4R_0^2}&0&-\frac{H\mathcal{F}^1_4(k)}{4R_0^2}&0&0&\vdots\\
\vdots&0&-\frac{H\mathcal{F}^1_6(k+1)}{4R_0^2}&0&\mathcal{S}^1_2(k+1)-\frac{H\mathcal{F}^1_2(k+1)}{4R_0^2}&0&-\frac{H\mathcal{F}^1_4(k+1)}{4R_0^2}&0&\vdots\\
\cdots&\cdots&\cdots&\cdots&\cdots&\cdots&\cdots&\cdots\\
\end{smallmatrix}\right]
\end{split}
\end{align}
similarly the first equation of~\eqref{e:3_2} is written as
\begin{equation}\label{e:3_7}
\mathcal{I}\begin{Bmatrix}
\ddot{a}_k
\end{Bmatrix}+A\begin{Bmatrix}
a_k
\end{Bmatrix}+B\begin{Bmatrix}
b_k
\end{Bmatrix}+A_1\begin{Bmatrix}
\dot{a}_k
\end{Bmatrix}+B_1\begin{Bmatrix}
\dot{b}_k
\end{Bmatrix}+\begin{Bmatrix}
-\frac{H}{2R_0}\\
\vdots\\
\end{Bmatrix}=0.
\end{equation}
where $\mathcal{I}$ is identity matrix, $A$ and $B$ are matrix obtained by linearizing $R$ and ignoring transients in $R$ as follows,
\begin{align}\label{e:3_8}
\begin{split}
A=
\left[\begin{smallmatrix}
\cdots&\cdots&\cdots&\cdots&\cdots&\cdots&\cdots&\cdots\\
\vdots&-\frac{H\mathcal{F}_5(k)}{4R_0^2}&0&\mathcal{S}_1(k)-\frac{H\mathcal{F}_1(k)}{4R_0^2}&0&-\frac{H\mathcal{F}_3(k)}{4R_0^2}&0&0&\vdots\\
\vdots&0&-\frac{H\mathcal{F}_5(k+1)}{4R_0^2}&0&\mathcal{S}_1(k+1)-\frac{H\mathcal{F}_1(k+1)}{4R_0^2}&0&-\frac{H\mathcal{F}_3(k+1)}{4R_0^2}&0&\vdots\\
\cdots&\cdots&\cdots&\cdots&\cdots&\cdots&\cdots&\cdots\\
\end{smallmatrix}\right],
\end{split}
\end{align}
\begin{align}\label{e:3_9}
\begin{split}
B=
\left[\begin{smallmatrix}
\cdots&\cdots&\cdots&\cdots&\cdots&\cdots&\cdots&\cdots\\
\vdots&-\frac{H\mathcal{F}_6(k)}{4R_0^2}&0&\mathcal{S}_2(k)-\frac{H\mathcal{F}_2(k)}{4R_0^2}&0&-\frac{H\mathcal{F}_4(k)}{4R_0^2}&0&0&\vdots\\
\vdots&0&-\frac{H\mathcal{F}_6(k+1)}{4R_0^2}&0&\mathcal{S}_2(k+1)-\frac{H\mathcal{F}_2(k+1)}{4R_0^2}&0&-\frac{H\mathcal{F}_4(k+1)}{4R_0^2}&0&\vdots\\
\cdots&\cdots&\cdots&\cdots&\cdots&\cdots&\cdots&\cdots\\
\end{smallmatrix}\right].
\end{split}
\end{align}
The expression for natural frequency becomes,
\begin{equation}\label{e:3_10}
\begin{Bmatrix}
\omega_k^2
\end{Bmatrix}={\rm spec}\begin{Bmatrix}
A-BB_a^{-1}A_a
\end{Bmatrix},
\end{equation}
where `spec' stands for the eigenvalues of a matrix. 
\section{Stability analysis}\label{s:4}
As the bubble oscillation is inherently non-spherical, the instability considered here is the change in nature of oscillatory behavior of shape mode $a_k$ from stable sinusoidal oscillations to exponential blowup in finite time as the pressure amplitude is increased. This phenomenon is referred to as exponential instability. Analyzing the stability of~\eqref{e:2_62}, \eqref{e:2_64} and \eqref{e:2_65} is in general very difficult as they are inhomogeneous coupled system of nonlinear ODEs. There are no general approaches to understand the stability of inhomogeneous equations. Considering the system of linear coupled equations
\begin{equation}\label{e:4_1}
\dot{\bs x}=\msb A(t)\bs x+\bs f(t),
\end{equation}
where $\bs x$ is $n$ dimensional vector and $\msb A(t)$ is $n\times n$ matrix. If the time period of both $\msb A(t)$ and $\bs f(t)$ are same, 
\begin{align}\label{e:4_2}
\begin{split}
\msb A(t)&=\msb A(t+T),\\
\bs f(t)&=\bs f(t+T),
\end{split}
\end{align}
then from the theorem proved in \citet{slane2011analysis} the exponential stability of \eqref{e:4_1} depends only on fundamental matrix $\msb X(T)$ of homogeneous part of the system of equations. The fundamental matrix can be constructed by integrating the system of equations~\eqref{e:4_1} till time period $T$ with initial conditions $\msb X(0)=\msb I$. The eigenvalues of the fundamental matrix determine the stability as follows,
\begin{itemize}
    \item Case 1: If all the eigenvalues are less than one then the system of equations is asymptotically stable. 
    \item Case 2: If at least one eigenvalue is greater than one then the system of equations is asymptotically unstable.
    \item Case 3: If some eigenvalues are equal to one and the remaining is less than one, then the solution is stable but not asymptotically stable.
\end{itemize}

The system of equations for radial and tangential shape modes is not of the form~\eqref{e:4_1}. However, substituting for $\ddot{R}$ in \eqref{e:2_64} from \eqref{e:2_62} and linearize radius as,
\begin{equation*}
R(t)=R_0(1+r(t)),
\end{equation*}
by ignoring $ra_k$, $r\dot{a}_k$, $rb_k$ and $r\dot{b}_k$ as small quantities, the system of equations of bubble can be rewritten in the form of~\eqref{e:4_1}. Now the only time-dependent term in the coefficient matrix is $p_\infty$ which is also the only time-dependent forcing term and if they are periodic then the above stability analysis holds. 
\section{Numerical results and discussion}\label{s:5}
The equations \eqref{e:2_62}, \eqref{e:2_64}, and \eqref{e:2_65}, are solved using the Runge-Kutta method implemented by MATLAB ode45 package, while equation~\eqref{e:2_52} is solved using the finite difference method. First equation~\eqref{e:2_52} is transformed as $y=r/R(t)$ and $\tau=t$ as
\begin{equation}\label{e:5_1}
\frac{\partial T_k}{\partial\tau}-\frac{\eta}{\rho R^2}\frac{\partial^2T_k}{\partial y^2}-\frac{\dot{R}}{R}(y-y^{-2})\frac{\partial T_k}{\partial y}+\left[\frac{\eta k(k+1)}{\rho R^2y^2}-\frac{2\dot{R}}{y^3R}\right]T_k=0,
\end{equation}
to have a fixed computational domain $[1,\infty]$ instead of $[R,\infty]$. Following~\citep{liu2012surface}, the convection term is discretized using a first-order upwind scheme and the diffusion equation using a second order central difference scheme as
\begin{align}\label{e:5_2}
\begin{split}
    \frac{\dot{R}(y-y^{-2})}{R}\frac{\partial T_k}{\partial r}&\approx\frac{\dot{R}(y-y^{-2})}{R}\frac{T^t_{k,i+1}-T^t_{k,i-1}}{2\Delta y}+\frac{|\dot{R}|(y-y^{-2})}{R}\frac{T^t_{k,i+1}-2T^t_{k,i}+T^t_{k,i-1}}{2\Delta y},\\
    \frac{\partial^2T_k}{\partial r^2}&\approx\frac{T^t_{k,i+1}-2T^t_{k,i}+T^t_{k,i-1}}{\Delta y^2}.
\end{split}
\end{align}
The sign of coefficients before convection term is only determined by $\dot{R}$ as $y\geq1$ and $R>0$. Using this equation~\eqref{e:5_1} gets discretized as
\begin{align}\label{e:5_3}
\begin{split}
&\frac{T^t_{k,i}-T^{t-1}_{k,i}}{\Delta t}-\left[\frac{\eta}{\rho R^2}+\frac{|\dot{R}|(y-y^{-2})\Delta y}{2R}\right]\frac{T^t_{k,i+1}-2T^t_{k,i}+T^t_{k,i-1}}{\Delta y^2}\\
&-\frac{\dot{R}(y-y^{-2})}{R}\frac{T^t_{k,i+1}-T^t_{k,i-1}}{2\Delta y}+\left[\frac{\eta k(k+1)}{\rho R^2y^2}-\frac{2\dot{R}}{y^3R}\right]T^t_{k,i}=0,
\end{split}
\end{align}
which can be solved to evaluate $T_k$. The problem is approached using a loosely coupled staggered scheme similar to~\citep{geuzaine2004second}. At each time step $ t_i $, the Runge-Kutta method is used to solve the mode equations, with the value of $ T_k $ at the previous time step $ t_{i-1} $ being utilized. This gives the solution for the variables of interest at time $ t $. Subsequently, the values of $ R $, $ a_k $, and $ b_k $ at time $ t_i $ are used to solve the finite difference equation, which updates the fluid terms for the next time step. The initial conditions are $R=R_0,a_k=0$ and $b_k=0$ for simulations. 
\subsection{Solution verification}
\begin{figure}
    \centering
    \includegraphics[width=1\linewidth]{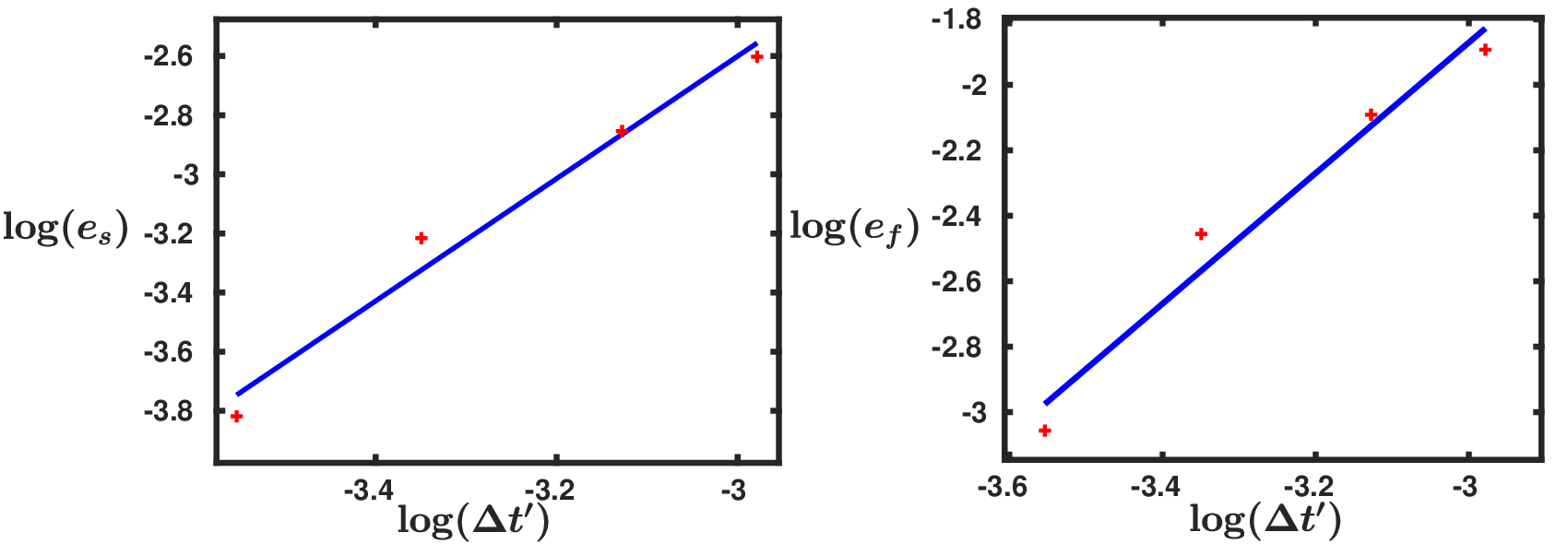}
    \caption{Error convergence plot for (a) Structural subsystem (b) Fluid subsystem}
    \label{fig:3}
\end{figure}
\begin{figure}
    \centering
    \includegraphics[width=1\linewidth]{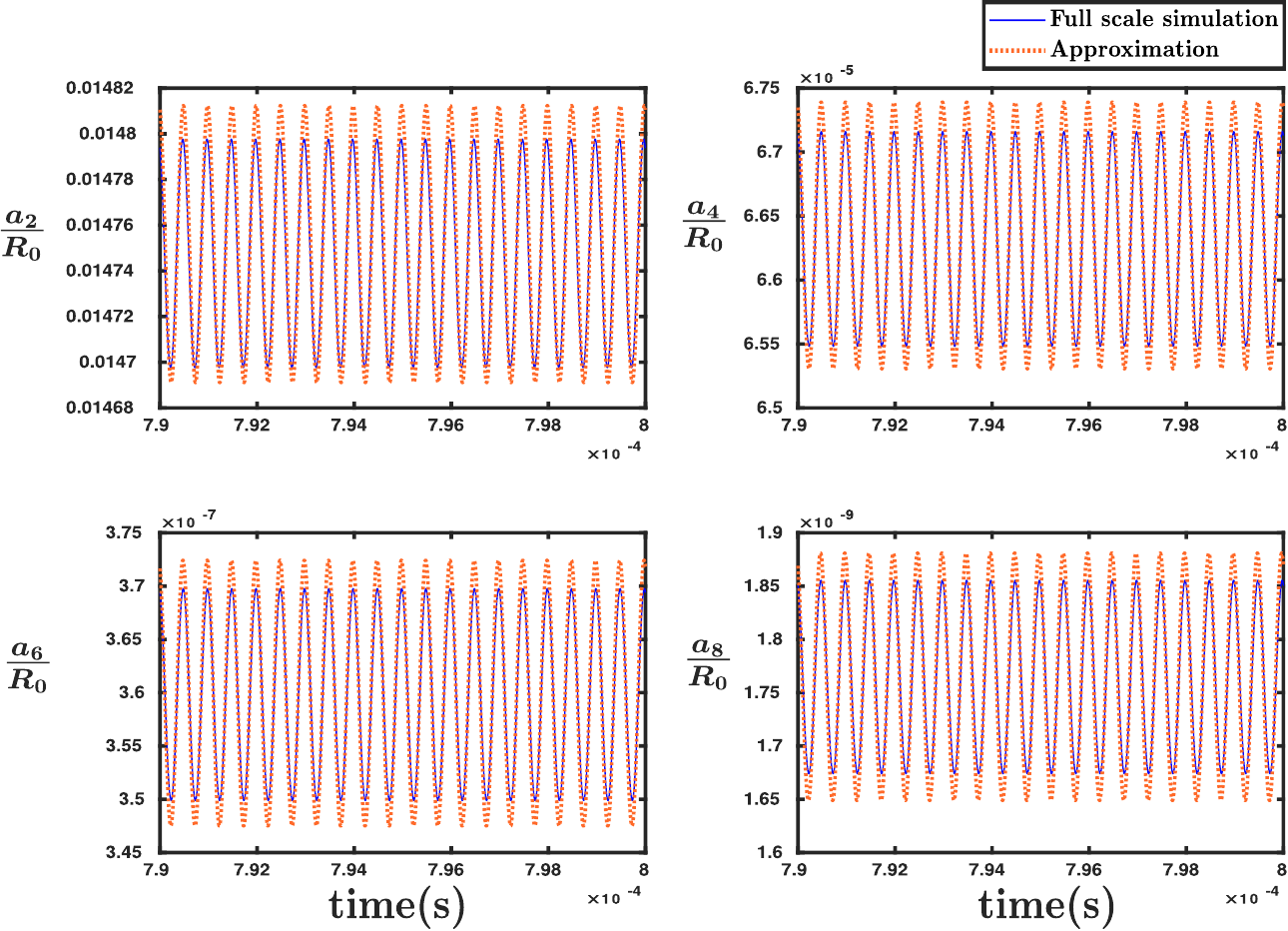}
    \caption{Comparison of full scale simulation with solution obtained by boundary layer approximation.}
    \label{fig:4}
\end{figure}
To prove the convergence of the numerical scheme as no exact solution to the system of equations is available a priori, a solution obtained with a very fine time step $\bar{a}_i(t')$ is used as the reference \citep{birken2010time,leveque2007finite}. The solutions computed with coarser time steps are then compared against this reference data to demonstrate convergence. As the magnitudes of the mode shapes lie in the range of $\mathcal{O}(10^{-3} - 10^{-30})$, only the relative error is considered. The system of equations has a fluid subsystem and a structural subsystem. Although the numerical algorithms for structure and fluid are order four and order one, the resulting loosely coupled method is neither fourth-order nor first-order accurate. To calculate the order of accuracy of the stagger scheme, the following error measures are introduced,
\begin{align}\label{e:5_4}
\begin{split}
e_s=\,\underset{i}{\rm max}\,\frac{\norm{a_i(t')-\bar{a}_i(t')}_2}{\norm{\bar{a}_i(t')}_2},\quad\quad
e_f=\,\underset{i}{\rm max}\,\frac{\norm{T^{100}_i-\bar{T}^{100}_i}_2}{\norm{\bar{T}^{100}_i}_2}.
\end{split}
\end{align}
Here, $\norm{.}_2$ denotes the $2$-norm, $T_i^{100}$ is $i^{\rm{th}}$ toroidal vortex at $100\,\rm{\mu s}$ and $a_i$ is $i^{\rm{th}}$ mode amplitude. To test the implementation, simulations are carried out for a total duration of $100\,\mu\text{s}$, with the reference solution $\bar{a}_i$ obtained using a non-dimensional time step of $\Delta t' = 1.7 \times 10^{-4}$. The convergence behavior is then assessed through log-log plots shown in figure~\ref{fig:3} where $\log$ refers to the common logarithm. It is observed that all the shape modes have qualitatively similar log-log plots. As expected, the maximum error is observed at higher mode numbers considered in the numerical analysis. This convergence in error proves that the time series converges to some solution. To ensure the correctness of the solution, the full-scale simulation is compared with the solution obtained by using the boundary layer approximation, as shown in figure~\ref {fig:4}. It is observed that the oscillation patterns and mean mode amplitudes are in close agreement, with only a small deviation in the maximum mode amplitude between the boundary layer approximation and the full-scale simulation. Based on these two results, it can be reasonably concluded that the numerical solution method provides correct results. Since the upwind scheme is used for discretizing the convection term, it must satisfy the CFL condition $(u_{\rm conv}\Delta t/\Delta y\leq1)$. Because the scheme is only conditionally stable, the solution becomes numerically unstable and eventually diverges when larger time steps $\Delta t'$ are used. The measured slope of $\log{e_s}$ plot is $1.995$, and $\log{e_f}$ plot is $2.065$, based on which it can be concluded that the loosely coupled scheme is approximately second-order accurate. From the error plots shown in figure~\ref{fig:3}, $\Delta t' = 1 \times 10^{-3}$ is chosen for our simulations. 
\subsection{Stability diagrams}
\begin{figure}
    \centering
    \includegraphics[width=1\linewidth]{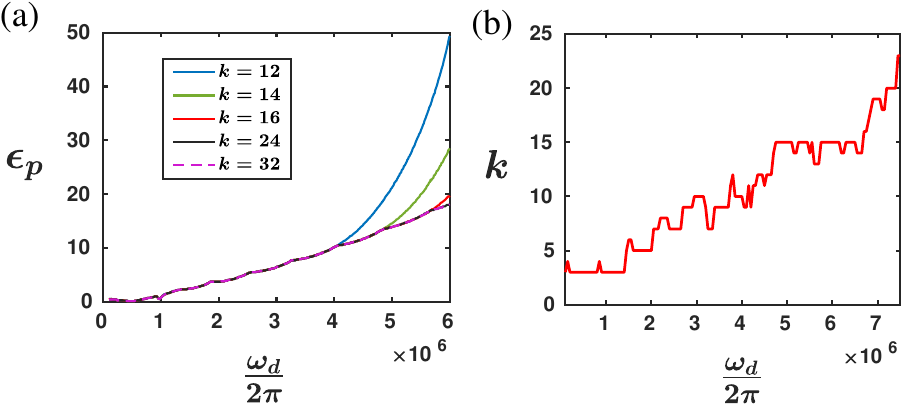}
    \caption{Convergence of stability diagram with number of modes and results of linear stability analysis. (a) Stability diagram with total modes considered in the analysis and (b) Mode number unstable at a particular frequency.}
    \label{fig:5}
\end{figure}

In the study of bubble dynamics under an applied magnetic field, all even-numbered shape modes are excited, making it challenging to determine the number of modes needed to be considered for analysis. To address this challenge, the pressure-frequency stability diagram (see figure~\ref{fig:5}a) is analyzed as a function of the number of modes, while applying pressure at infinity as $p_{\infty}=p_0(1+\epsilon_p\cos{\omega_dt}).$ Here, $\epsilon_p$ and $\omega_d$ represent dimensionless acoustic pressure with ambient pressure and circular frequency, respectively. The stability curve represents the critical pressure above which the amplitude response of at least one nonspherical mode $a_k$ changes its behavior from bounded oscillations to exponential blowup.
In the simulations, the coils are assumed to carry a current of  $I = 100$ kA, with the number of coil turns set to $ N_1 = 1000$, a thickness of $\epsilon = 20 $ nm, and an initial bubble radius of $R_0 = 10$ $\mu$m.  

Figure~\ref{fig:5}a shows that as the number of modes increases from 12 to 32, the critical pressure curves in the $\epsilon_p - \omega_d $ plane stabilize, defining a well-converged stable region where the linearized non-spherical oscillation equations remain valid. This convergence occurs because, in the absence of initial disturbances in the shape modes, the $a_2$ mode is initially dominant. The amplitude of each successive even-numbered mode gradually decreases, approaching nearly zero at higher mode numbers.  

Furthermore, the stability analysis results discussed in Section \ref{s:4} indicate that the first mode to become unstable during a frequency sweep shifts to higher modes as frequency increases, as shown in figure~\ref{fig:5}b. This explains the convergence of the critical pressure curve when more than 16 modes are considered for driving frequencies up to $6$ MHz.
\begin{figure}
    \centering
    \includegraphics[width=1\linewidth]{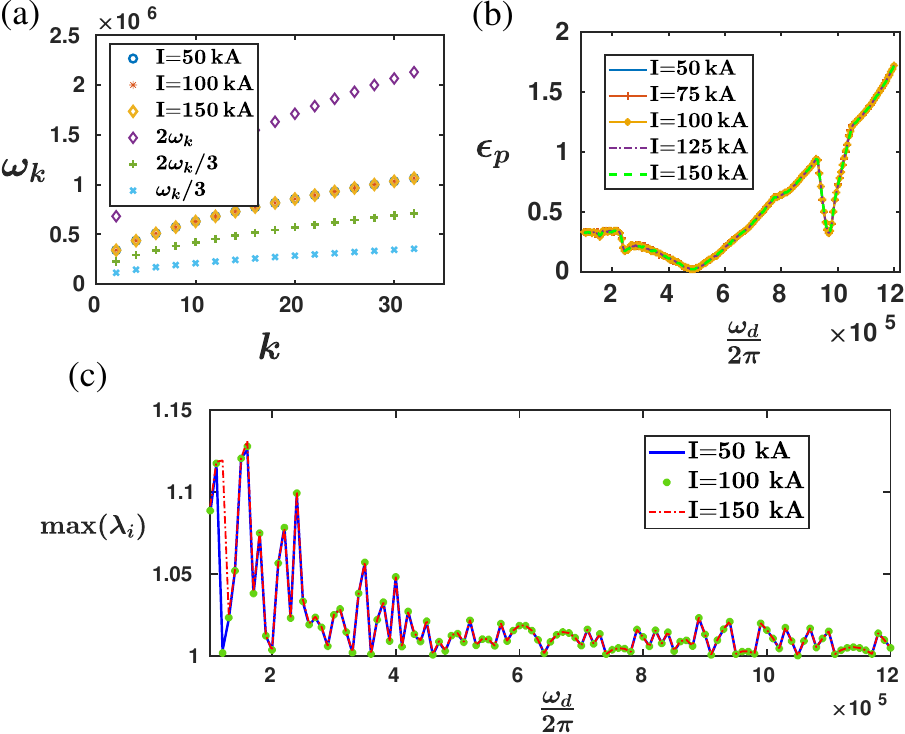}
    \caption{(a) Variation of natural frequency with mode numbers plot $\omega_k\,\,\rm{vs}\,\,k$ at various values of applied static current, (b) The critical pressure boundary in the pressure frequency plane $\epsilon_p\,\,\rm{vs}\,\,\omega_d$ shown for different values of applied static current   (c) Maximum eigenvalue of fundamental matrix variation with driving frequency at various values of applied current.}
    \label{fig:6}
\end{figure}
\subsubsection{Effect of static current and material parameters on stability diagram for coil field}

Figure~\ref{fig:6}b shows the influence of applied static current on the critical pressure above which mode amplitude response becomes unstable. From this, it is evident that the applied magnetic field does not affect the pressure-frequency stability diagram of surface instability. Similarly, the $\omega_k - k$ plot in figure~\ref{fig:6}a, which shows the relation between natural frequency and mode number, suggests that the applied current does not noticeably influence the natural frequency. Additionally, the same plot reveals that several harmonics, sub-harmonics, and super-harmonics of different modes are closely spaced around the fundamental frequency of a particular mode. As a result, when the critical pressure is exceeded, multiple modes become unstable simultaneously. This complexity makes stability analysis feasible only through computational methods.
\begin{figure}
    \centering
    \includegraphics[width=1\linewidth,trim=0in 0in 0in 0in]{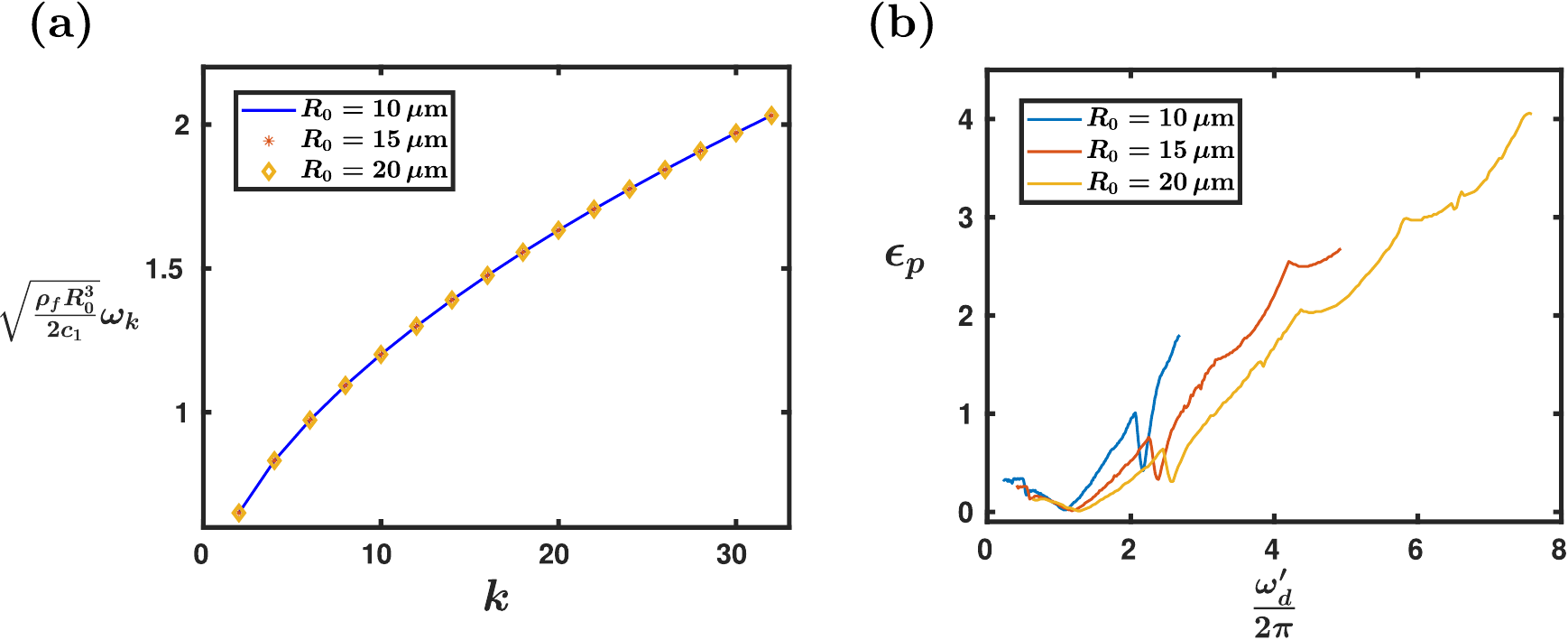}
    \caption{(a) Variation of natural frequency with mode numbers for shear modulus $c_1=0.1\,{\rm N\,m}^{-1}$ and (b) critical pressure versus driving frequency curve for shear modulus $c_1=0.1\,{\rm N\,m}^{-1}$ in $\epsilon_p\,\,{\rm vs}\,\,\omega_d'$ plane shown for different values of initial radius.}
    \label{fig:7}
\end{figure}
\begin{figure}
    \centering
    \includegraphics[width=1\linewidth]{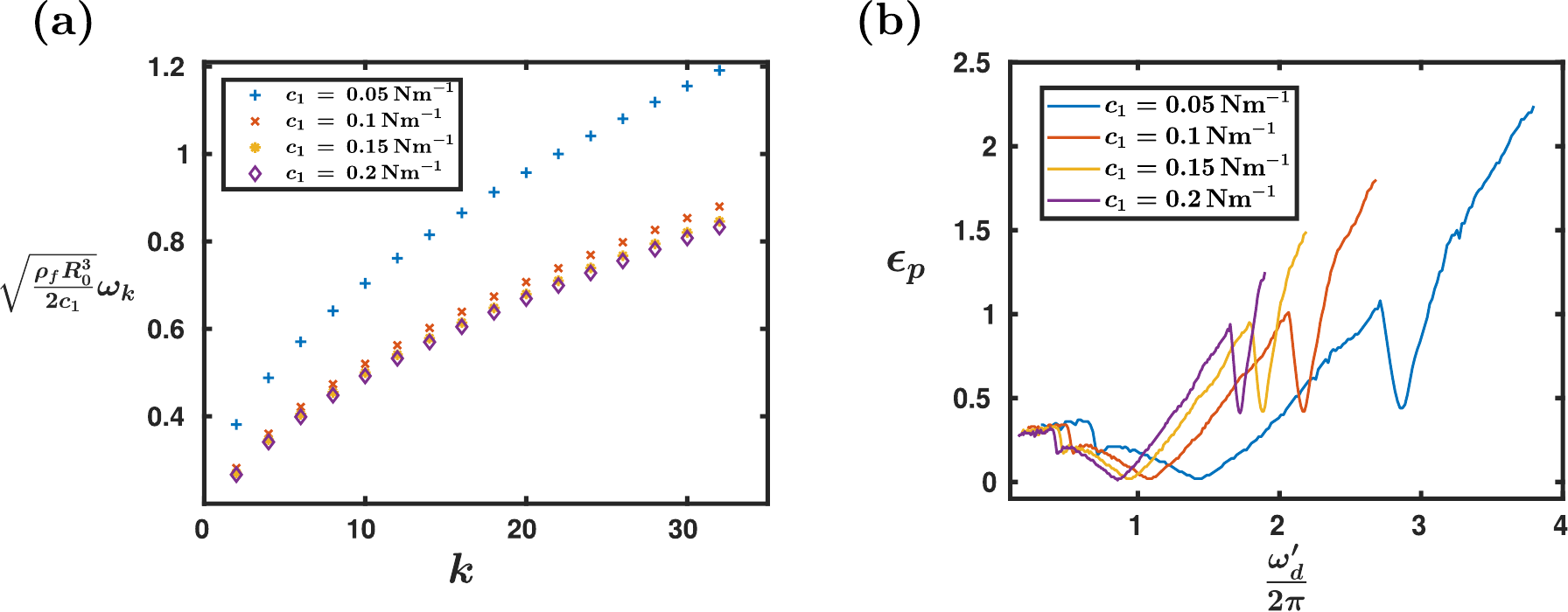}
    \caption{Plots for an undeformed bubble radius $R_0=10\,\mu\rm{m}$ with different values of surface encapsulation shear modulus $c_1$. (a) Variation of natural frequency with mode numbers. (b) Critical excitation pressure ($\epsilon_p$) versus driving frequency ($\omega_d$) curves representing the stability boundaries.}
    \label{fig:8}
\end{figure}
Figure~\ref{fig:6}c is obtained by calculating the maximum eigenvalue of the fundamental matrix $X(T)$ by varying the pressure amplitude until the eigenvalue becomes greater than one. The $\text{max}(\lambda_i)$ plot of figure~\ref{fig:4}c indicates that the maximum eigenvalue remains unaffected by increasing current in the system of equations, which explains the invariance of the stability diagram with respect to variations in current. Furthermore, frequency is non-dimensionalised as $\omega'=\sqrt{\rho_f R_0^3/2c_1}$ and ambient pressure as $P=p_0R_0/c_1$. As shown in figure~\ref{fig:7}a, this scaling causes the frequency curves to collapse onto a single curve when the undeformed radius $R_0$ is varied. On the other hand, the frequency curves do not collapse when $c_1$ is varied, as seen in figure~\ref{fig:8}a. This is due to the presence of $c_2$ term in the Mooney-Rivlin model, which introduces an additional non-dimensional parameter $\alpha={c_2}/{c_1}$ that affects the scaling. As $c_1$ increases while $c_2$ remains fixed, $\alpha$ decreases, reducing its influence and causing the frequency curves to converge.
The pressure–frequency stability curves do not collapse when either $R_0$ or $c_1$ is varied. This is because the instability with increasing pressure arises from the presence of the $\ddot{R}$ term in equation~\eqref{e:2_64}. Substituting $\ddot{R}$ from equation~\eqref{e:2_62} and simplifying leads to a form resembling Mathieu’s equation,
$\ddot{q}+(\delta+\eta \cos{\omega_d t})q=0$,
with some extra terms. The stability of this modified equation is highly sensitive to parameters $\delta$ and $\eta$, which depend on $R_0$, $c_1$ and $\omega_d$. So, stability curves do not collapse to a single curve as the undeformed radius or surface shear modulus is varied.
\subsubsection{Effect of time varying current in the coil on stability}
\begin{figure}
    \centering
    \includegraphics[width=0.55\linewidth]{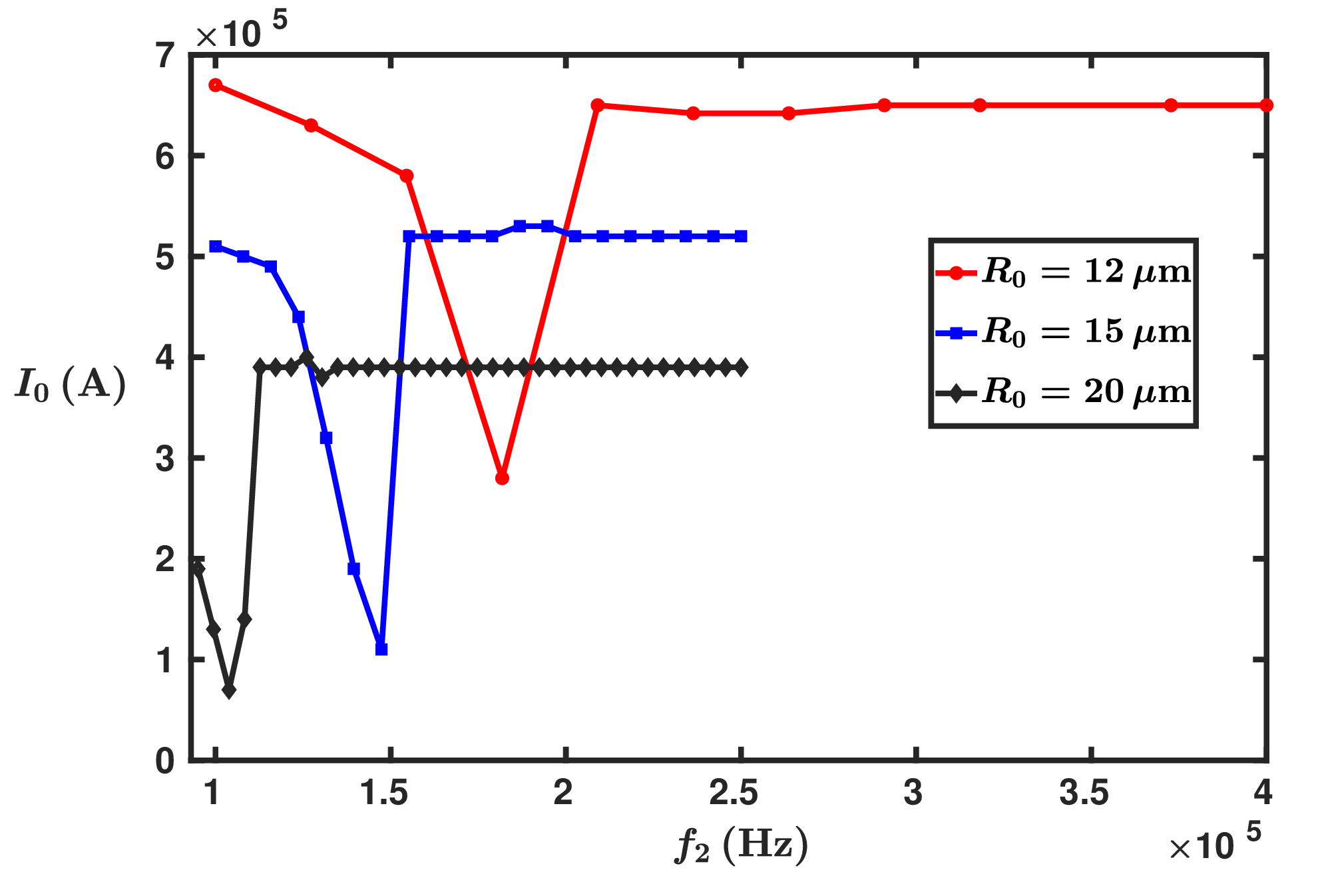}
    \caption{Critical current amplitude\,($I_0$) versus frequency\,$(f_2)$ with $\epsilon_p=0$, $c_1=0.1\,{\rm N\,m}^{-1}$ for different values of initial radius.}
    \label{fig:9}
\end{figure}

Time-varying current and pressure are employed as,
\begin{align}\label{e:5_5}
p_\infty=p_0(1+\epsilon_p\cos{\omega_1t}),\quad
I=I_0\cos{\omega_2t}.
\end{align}
Here, $I_0$ denotes the dimensional current. When only a time-varying current is applied, with $\epsilon_p = 0$ and $\omega_2 = 2\pi f_2$, the resulting current–frequency stability diagram is obtained for various radii, as shown in figure~\ref{fig:9}. In this study the magnetic field of range $0.5-1.5\,{\rm T}$ is used as it is the range used in medical application and initial radii of bubble considered is $10-20\,\mu{\rm m}$, from this the maximum current amplitude lies in the range $133-266\,{\rm kA}$. From the plots, it can be seen that instability occurs mainly at higher current amplitudes, except at a few frequencies. At higher frequencies, no dips are observed, which appear only at lower frequencies. Since the magnetic force is proportional to $H_2R_0^2$, increasing the radius reduces the critical current amplitude. 
\begin{figure}
    \centering
    \includegraphics[width=1\linewidth]{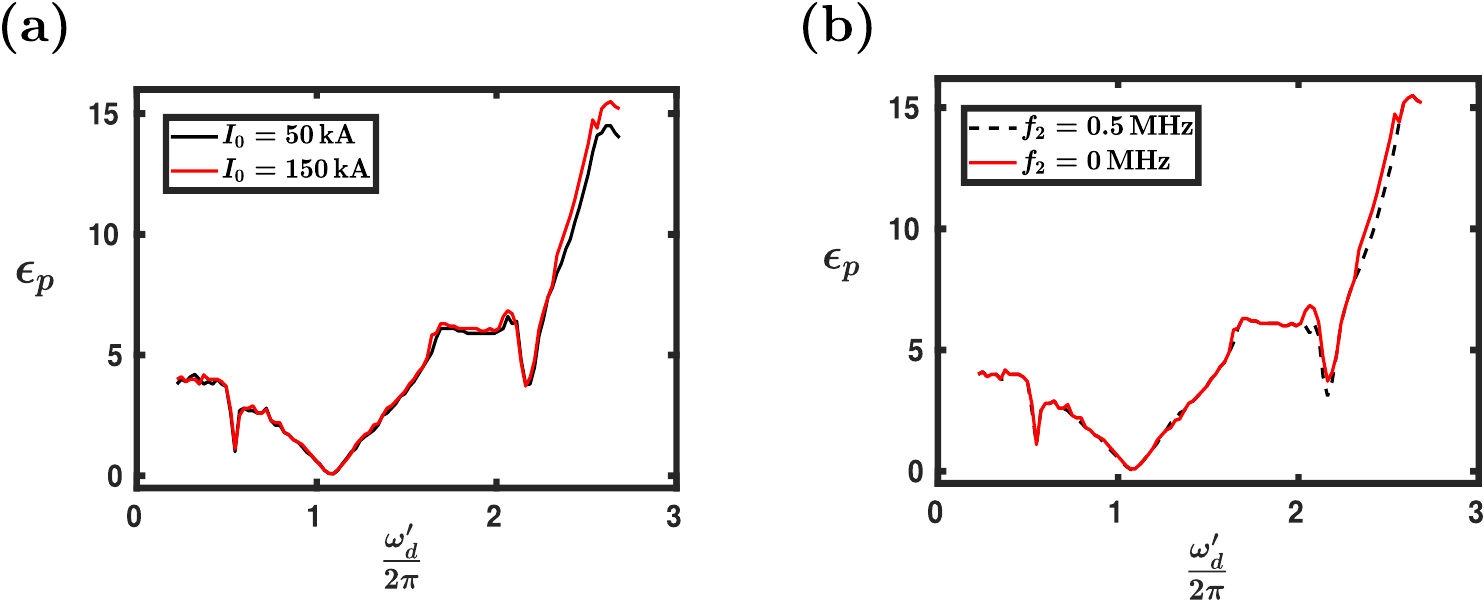}
\caption{Stability diagram for time varying current (a) Critical pressure amplitude versus frequency curves in $\epsilon_p$ vs $\omega'_d$ plane for $f_2=0.5\,{\rm MHz}$, initial radius $R_0=10\,\mu{\rm m}$ and $c_1=0.1\,{\rm N\,m}^{-1}$ (b) Critical pressure amplitude versus frequency curves in $\epsilon_p$ vs $f_d$ plane for $I_0=150\,{\rm kA}$, initial radius $R_0=10\,\mu{\rm m}$ and $c_1=0.1\,{\rm N\,m}^{-1}$ for different frequencies.}
    \label{fig:10}
\end{figure}

From the pressure–frequency stability diagram shown in figure~\ref{fig:10}, it can be observed that the frequency and amplitude of the applied current have little effect on the stability plot, with noticeable variation appearing only at higher pressure frequencies $f_d$. It is also observed from other simulations that, for other frequencies and amplitudes of current, the critical pressure curves lie within these two bounds. Interpreting the simulation results becomes challenging, since the governing equations now take the form of a dual-frequency Mathieu’s equation
\begin{equation}\label{e:5_6}
    \ddot{x}+(\delta+\epsilon_1\cos{\omega_1t}+\epsilon_2\cos{\omega_2t})x=0,
\end{equation}
with nonlinear coefficients. A separate, detailed analysis using perturbation methods is needed to gain deeper insights into bubble behavior. 

Another way of generating a time-varying magnetic field is to use oscillating coils with distance $a$ (shown in figure~\ref{fig:2}) varying as $a=\tilde{a}(1+\beta \cos{2\pi f_3 t})$. The amplitude of coil oscillation is assumed to be small with $\beta\ll1.$ However, from the simulations, it is observed that neither $\beta$ nor $f_3$  influences pressure frequency stability plots (not presented in the results section).  On the other hand, the periodicity of the modal oscillations depends on the relation between $f_1$ and $f_3$, which is also observed in the case of time-varying current. Furthermore, the analysis presented here pertains to specific sources, considering that a more general applied magnetic field could offer a broader understanding of magnetic bubble dynamics.
\subsubsection{Stability diagrams for dipole field}
\begin{figure}
    \centering
    \includegraphics[width=\linewidth]{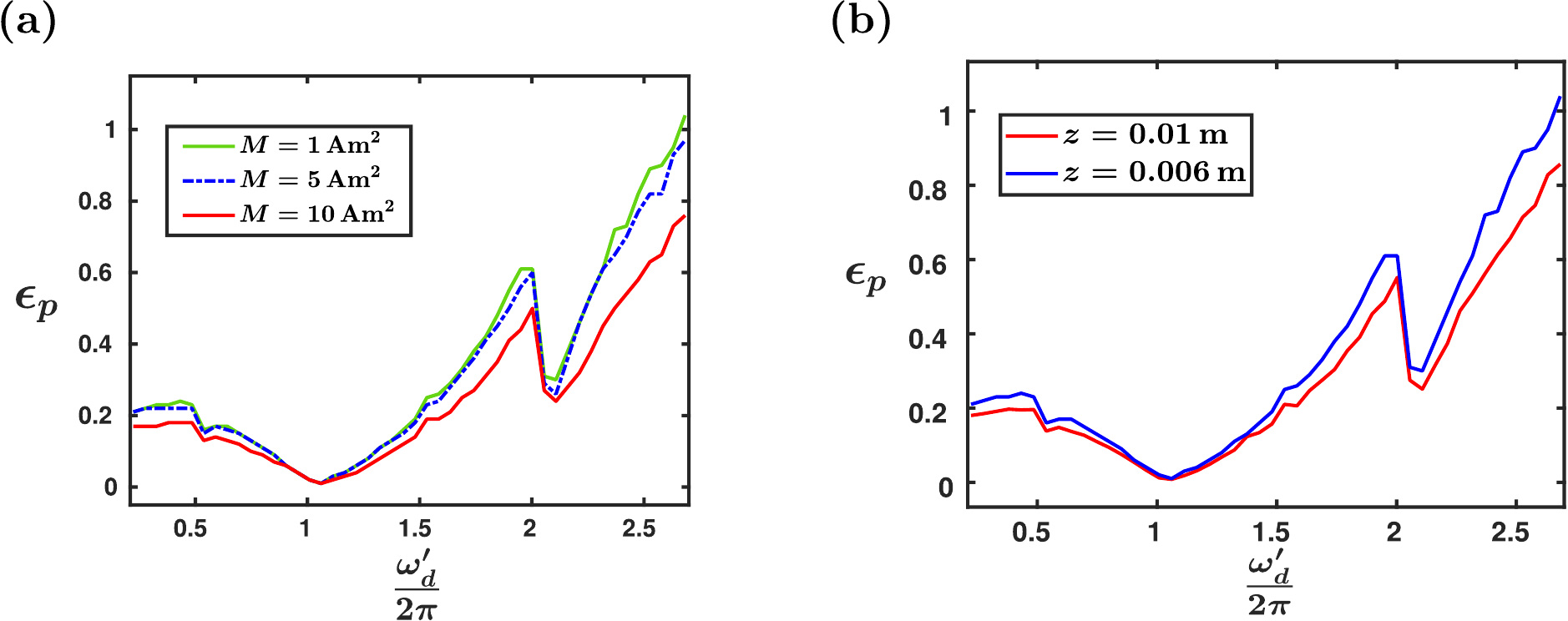}
    \caption{Critical pressure vs natural frequency curves for a bubble of shear modulus $c_1=0.1\,\rm{Nm}^{-1}$, radius $R_0=10\,\mu\rm{m}$ for various values of (a) dipole strength at $z=0.01\,\rm{m}$ and (b) distance of dipole from center of bubble at $M=1\,\rm{Am}^2$.}
    \label{fig:11}
\end{figure}
The computation is performed for bubble of radius $R_0=10\,\mu\rm{m}$, shear modulus $c_1=0.1\,\rm{Nm}^{-1}$ and susceptibility $\chi=5$. From figure~\ref{fig:11} it is observed that stability decreases as dipole strength $M$ is increased and the distance of the dipole from the center of the bubble is decreased. The non-dimensional terms for magnetic forcing is,
\begin{equation}\label{e:5_7}
\mathcal{M}_n=\frac{12\mu_0\chi M^2R_0^2}{z^8c_1}\sim \mathcal{O}(10^{-4}-10^{-2}).
\end{equation}
The above order of magnitude is for $z=0.01\,\rm{m}$. Although the order of magnitude of forcing is the same as that of the coil field, the dipole field slightly decreases $\delta$ compared to that of the coil field in Mathieu's equation $\ddot{q}+(\delta+\eta\cos{\omega_dt})q=0$, which makes it enter the unstable zone at a lower pressure amplitude, decreasing the stability. Here also at lower values of dipole strength, the stability diagrams remain nearly the same. Decreasing $z$ by half increases the order of forcing 200 times and significantly reduces the stability zone.
\subsection{Time-series analysis of mode shapes}
Radial and shape oscillations are analyzed only within the stable region of the $\epsilon_p-\omega_d$ plane. In this stable zone, the dynamics are predominantly governed by $R(t)$ and $a_2(t)$. The behavior of these two modes is studied as material properties vary. To investigate the influence of different material properties on the time series of radial and shape modes, the governing equations are non-dimensionalized. This is done by introducing the transformations: $$R(t)=R_0\bar{R}(t),\,a_k=R_0\bar{a}_k(t), \,t=\sqrt{\frac{\rho_fR_0^3}{2c_1}}\,\bar{t},$$ where $c_1=\epsilon\,C_1$ and $c_2=\epsilon\,C_2$. These transformations lead to the introduction of key non-dimensional quantities: Reynolds number $Re=\sqrt{{2\rho_fc_1R_0}/{\eta^2}}$, which compares elastic and viscous forces in liquid, non-dimensional magnetic force parameter $Rn=\frac{\mu_o\chi_s H_2^2R_0^2}{2c_1a^2}$ for coil field and $\mathcal{M}_n$ defined in equation~\eqref{e:5_7} for dipole field and strain stiffening parameter $\alpha={c_2}/{c_1}$. These non-dimensional parameters facilitate a systematic exploration of how material properties influence the dynamics of the system.
\begin{figure}
    \centering
    \includegraphics[width=1.1\linewidth]{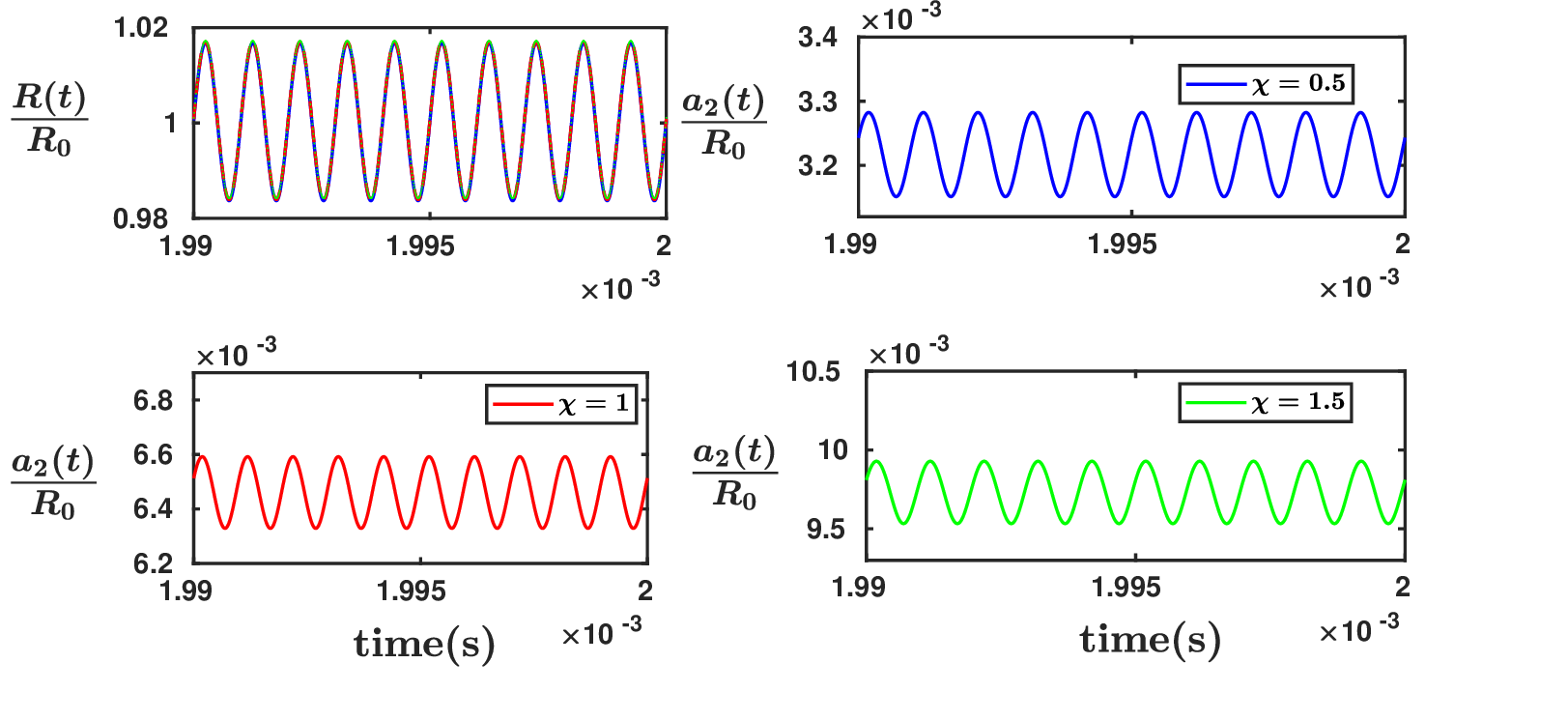}
    \caption{Variation of radial mode amplitude $R(t)$ and second mode amplitude $a_2(t)$ with time for a bubble of initial radius $R_0=10\,\mu\rm{m}$ at various values of encapsulation susceptibility. The simulations are conducted at $\omega_d=1\,\rm{MHz}$ and $\epsilon_p=0.5$.}
    \label{fig:12}
\end{figure}
\begin{figure}
    \centering
    \includegraphics[width=1\linewidth]{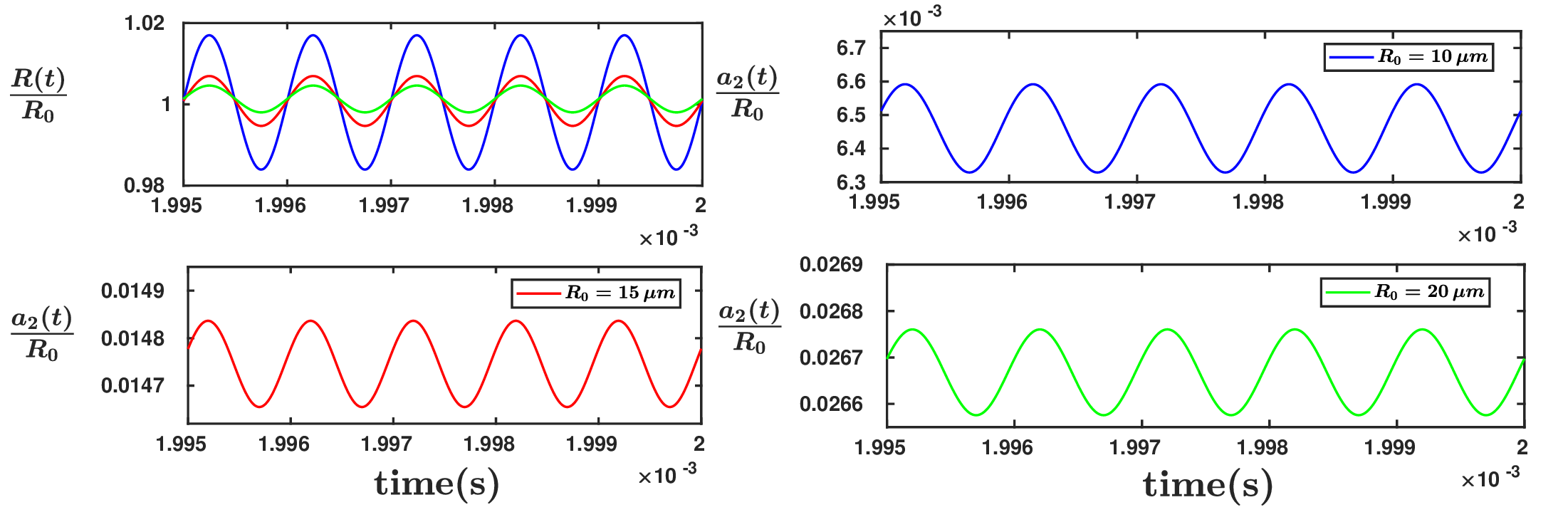}
    \caption{Variation of radial mode amplitude $R(t)$ and second mode amplitude $a_2(t)$ with time for a bubble encapsulation of susceptibility $\chi=1$ at different values of undeformed bubble radius $R_0$. The simulations are conducted at $\omega_d=1\,\rm{MHz}$ and $\epsilon_p=0.5$.}
    \label{fig:13}
\end{figure}
\subsubsection{Variation with magnetic susceptibility for coil field}
Simulations are conducted with the following parameters: $R_0=10\,\mu\rm{m}$, $c_1=0.1\,\rm{N\,m}^{-1}$, $c_2=0.12\,\rm{N\,m}^{-1}$, $N_1=1000$ turns, $\epsilon=20\rm\,{nm}$, $\omega_d=1\,\rm{MHz}$, $\epsilon_p=0.5$ and $I=100\,\rm{kA}$ over a duration of $2000\,\mu\rm{s}$. The time-dependent behavior of the radial mode and the $a_2$ shape mode is analyzed. The results show that the radial mode remains largely unaffected, which can be explained by the relative magnitudes of the non-dimensional forces involved. This is because
the order of magnitude of various terms are $R_0\sim\mathcal{O}(10^{-5})$, $c_1\sim\mathcal{O}(0.1)$, $\chi_s\sim\mathcal{O}(10^{-8})$, $H_2\sim\mathcal{O}(\frac{IN_1}{a})\sim\mathcal{O}(10^{9})$ and $p_\infty\sim\mathcal{O}(10^5)$ which gives order of magnitude of pressure force and magnetic force as
\begin{align}\label{e:5_8}
\begin{split}
P&=\frac{p_0 R_0}{c_1}\sim\mathcal{O}(10^1),\\
Rn&=\frac{\mu_o\chi_s H_2^2R_0^2}{2c_1a^2}\sim\mathcal{O}(10^{-2}-10^{-3}).
\end{split}
\end{align}
Therefore, the magnetic forces are significantly smaller compared to the applied pressure, rendering their influence negligible on increasing $\chi$. However, for the $a_2$ shape mode, magnetic force serves as the primary driving term, significantly influencing its maximum oscillation amplitude. This dependence on susceptibility is clearly observed, as illustrated in figure~\ref{fig:12}.
\begin{figure}
    \centering
    \includegraphics[width=1\linewidth]{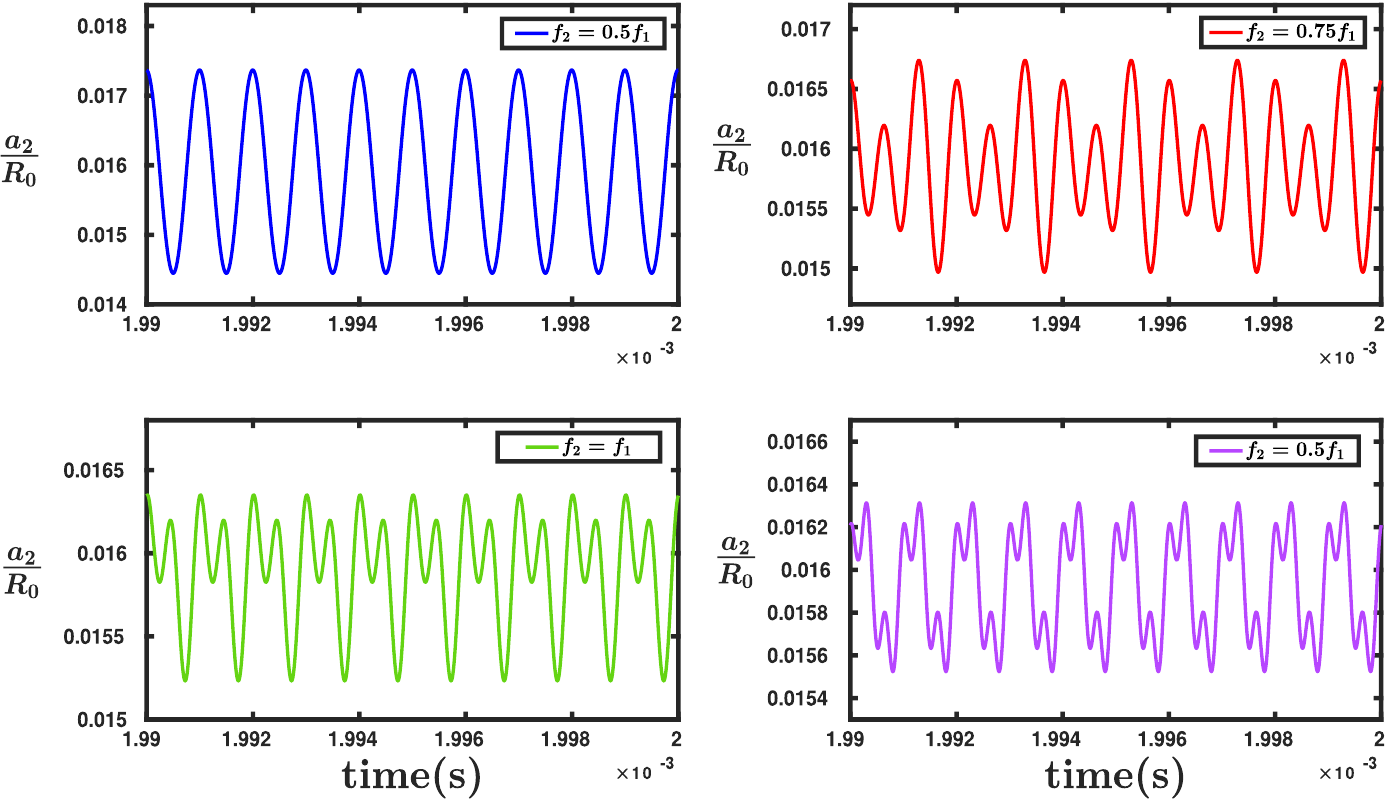}
    \caption{Variation of second mode amplitude $a_2(t)$ with time for a bubble encapsulation of susceptibility $\chi=1$, initial radius $R_0=10\,\mu{\rm m}$, current amplitude $I_0=150\,{\rm kA}$, pressure amplitude $\epsilon_p=0.5$ and frequency of pressure $f_1=1\,{\rm MHz}$ for different frequency of current $f_2$.}
    \label{fig:14}
\end{figure}
\subsubsection{Variation with undeformed radius for coil field}
Simulations are conducted with $c_1=0.1\,\rm{N\,m}^{-1}$, $c_2=0.12\,\rm{N\,m}^{-1}$, $\chi=1$, $\epsilon=20\,\rm{nm}$, $\omega_d=1\,\rm{MHz}$, $\epsilon_p=0.5$ and $I=100\,\rm{kA}$ over a duration of $100\,\mu\rm{s}$. From the expressions for the non-dimensional magnetic force, it is evident that the forcing on shape modes increases with an increase in radius. Additionally, as shown in \eqref{e:2_62}, the pressure acting on the radial mode opposes the radial motion, which leads to a reduction in the amplitude of radial oscillations with increasing radius. This behavior is clearly depicted in figure~\ref{fig:13}. Also, pressure force is proportional to $R_0$ and magnetic forcing is proportional to $R_0^2$. This scaling is reflected in the $a_2$ response, where its amplitude increases, respectively, by approximately $2.25$ and $4$ times as $R_0$ varies from $10-20\,\mu$m.
\subsubsection{Time series analysis of modes with applied frequency of current in the coil}
\begin{figure}
    \centering
    \includegraphics[width=1\linewidth]{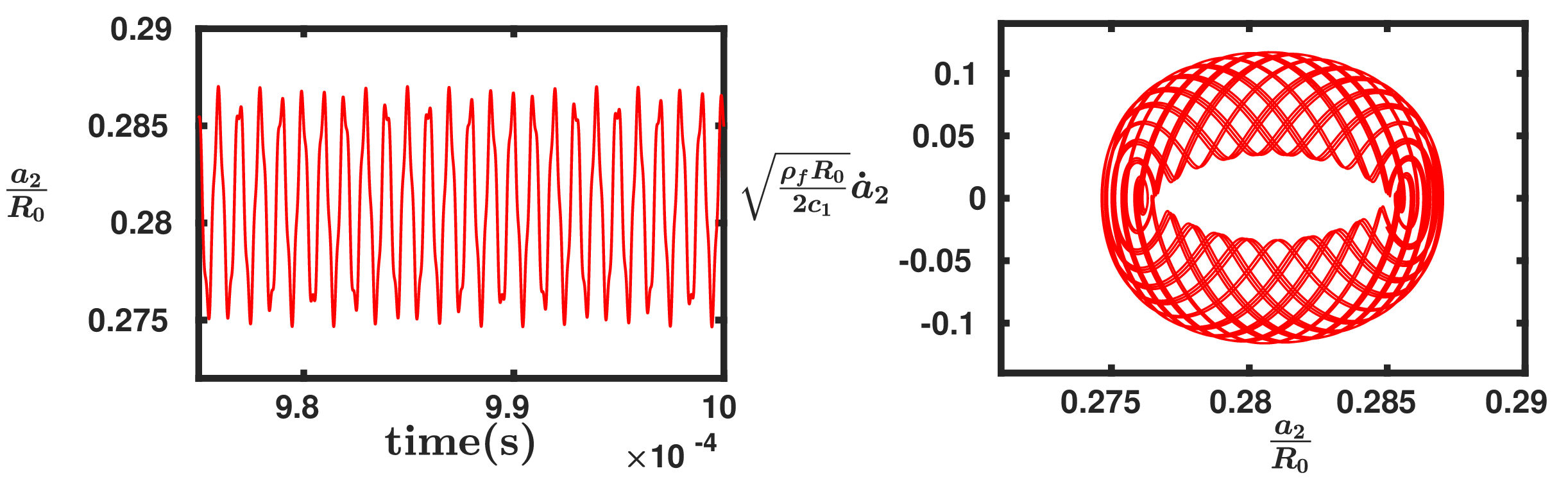}
    \caption{Variation of second mode amplitude $a_2(t)$ and phase plot with time for a bubble encapsulation of susceptibility $\chi=1$, initial radius $R_0=10\,\mu{\rm m}$, current amplitude $I_0=150\,{\rm kA}$, pressure amplitude $\epsilon_p=0.5$ and frequency of pressure $f_1=1\,{\rm MHz}$ and $f_2=1.777f_1$.}
    \label{fig:15}
\end{figure}
\begin{figure}
    \centering
    \includegraphics[width=1\linewidth]{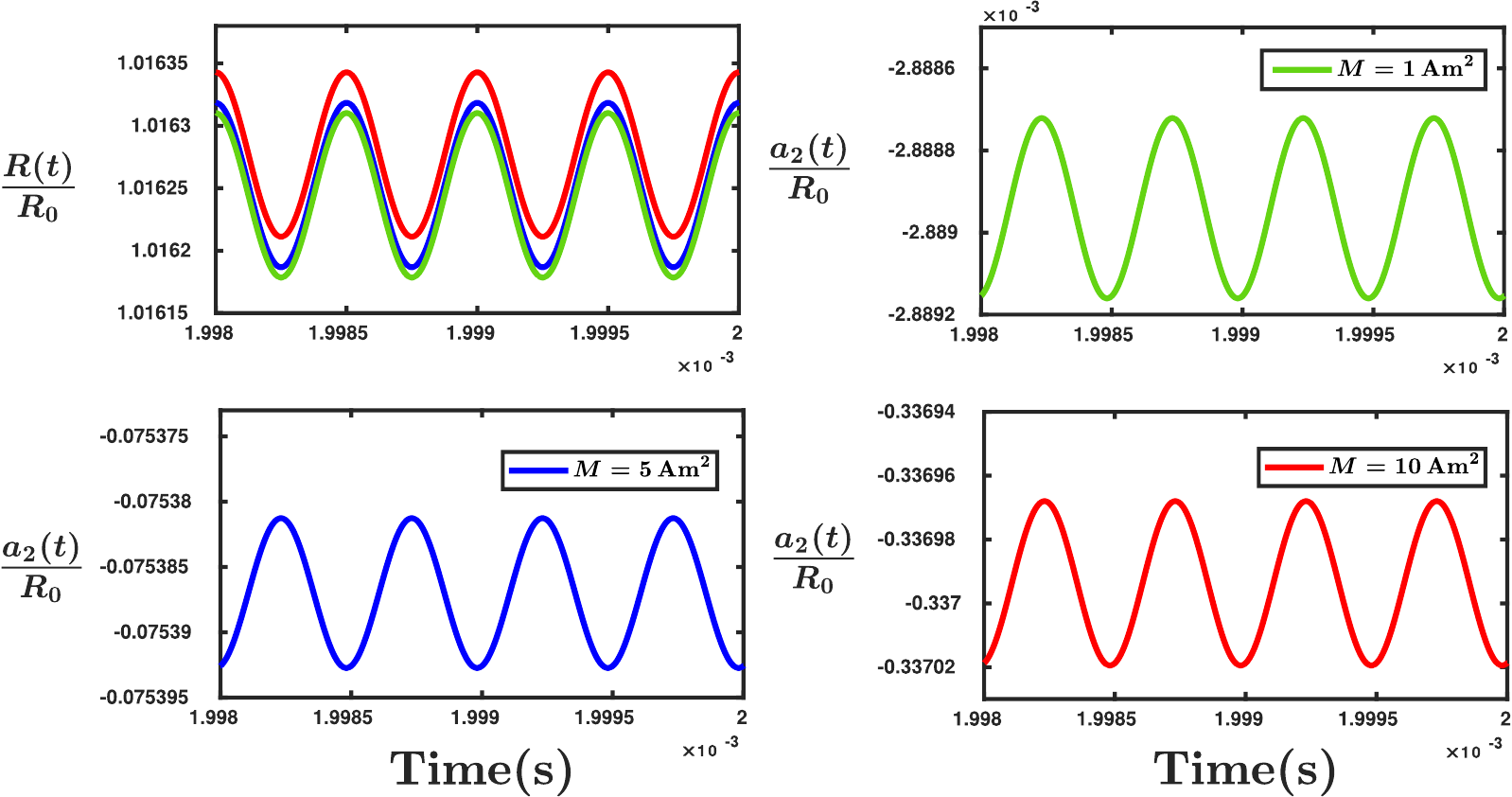}
    \caption{Time series response of radial second and fourth mode for a bubble of radius $R_0=10\,\mu\rm{m}$, surface shear modulus $c_1=0.1\,\rm{Nm}^{-1},$ susceptibility $\chi=5$ for different values of dipole strength $M$.}
    \label{fig:16}
\end{figure}
It can be seen from figure~\ref{fig:14} that as the frequency of applied current is increased, proportional to the frequency of the pressure field, the periodicity of the oscillation pattern of $a_2$ mode continuously varies with pressure and current varying as,
\begin{align}\label{e:5_9}
p_\infty=p_0(1+\epsilon_p\cos{\omega_1t}),
\quad I=I_0\cos{\omega_2t}.
\end{align}
Here, $\omega_1=2\pi f_1$ is the angular frequency of the pressure field and $\omega_2=2\pi f_2$ is the angular frequency of the applied current. The pattern of only the $a_2$ mode is studied, as it is the dominant mode within the stable zone. The frequency of the current is found to have only a slight influence on the amplitude of the $a_2$ mode, with $f_2 = 0.5f_1$ producing the maximum effect. This occurs because $\cos^2{\omega_2 t}$ expands to $1 + \cos{2\omega_2 t}$, which directly adds to the pressure force, thereby increasing the amplitude. Further simulations reveal that the oscillations are predominantly chaotic, except when $f_2$ is a fractional multiple of $f_1$, as shown in figure~\ref{fig:15}. This observation is noteworthy, since operating at frequencies that induce strong chaotic behavior makes controlling bubble dynamics increasingly challenging.  A more comprehensive analysis of bifurcation patterns under varying material parameters may be conducted separately to gain deeper insights into the oscillatory characteristics of magnetic bubbles. 
\subsubsection{Time series variation with dipole strength for dipole field}
Simulations are conducted with the following parameters: $R_0=10\,\mu\rm{m}$, $c_1=0.1\,\rm{N\,m}^{-1}$, $c_2=0.12\,\rm{N\,m}^{-1}$, $\omega_d=1\,\rm{MHz}$, $\chi=5$, $z=0.01\,\rm{m}$ and $\epsilon_p=0.5$ over a duration of $2000\,\mu\rm{s}$. It can be observed from figure~\ref{fig:16}, that increasing the dipole strength leads to a slight increase in the amplitude of radial oscillations, while significantly enhancing the magnitude of the second mode oscillation. From the expression for $\mathcal{M}_n$ in equation~\eqref{e:5_7}, it is evident that the magnetic forcing scales with $M^2$. Consequently, increasing the dipole strength by a factor of ten amplifies the magnetic forcing by a factor of one hundred, leading to a noticeable impact on radial oscillations and a pronounced effect on second mode.
\section{Summary and conclusions}\label{s:6}
In this paper, governing equations are derived for the non-spherical oscillations of an encapsulated magnetic microbubble using a membrane approximation of weakly magnetic hyperelastic materials. The mode shape oscillations are analyzed under the influence of time-varying pressure in the presence of two different external magnetic field configurations. In the first configuration, the magnetic field is generated by a pair of oppositely directed current-carrying coils, while in the second configuration, it is produced by symmetrically positioned magnetic dipoles. For the case of coil-field, numerical investigation has shown that the applied static current does not have a significant influence on the pressure-frequency stability, a conclusion further supported by stability analysis and order of magnitude analysis. Additionally, the time-varying current has only a minor influence on stability, as increasing the coil current amplitude makes the system exceed the linear region before the exponential blow-up of modes, thereby rendering the equations invalid. Under dipole field, high values of dipole strength, as well as decreasing the distance of dipole from the bubble center, significantly reduce the stability zone and make the bubble unstable. The applied static current does affect the amplitude of the second mode. Furthermore, increasing the undeformed bubble radius enhances the amplitude of the second mode while suppressing radial mode oscillations. Similarly, increasing the interface magnetic susceptibility amplifies the second mode but has little to no effect on radial oscillation amplitude for the coil field. The frequency of the time-varying current has only a minor effect on the amplitude of the mode shapes; however, it significantly influences their periodicity and dynamic response. Increasing the dipole strength slightly increases the amplitude of radial oscillation and significantly influences the amplitude of the second mode.

Using the viscous penetration depth assumption for a toroidal component of vorticity and an order-of-magnitude analysis, we formulate a method to calculate the natural frequency of shape modes for the coil field. Variations in material properties are shown to cause shifts in the stability diagram. But, natural frequency curves collapse to a single curve when the undeformed radius is varied, and it does not collapse when $c_1$ is varied, which is due to the presence of $c_2$, a material constant. However, the model has notable limitations. The membrane approximation is valid primarily for slightly larger bubbles; for smaller bubbles, a shell theory would provide a more accurate approximation. Furthermore, as we focus on linear shape mode oscillations, the derived equations are only valid within a limited stable region. Beyond this region, the equations exhibit exponential divergence. Incorporating non-linear terms for shape oscillations could potentially address this divergence and lead to stabilization. The work can be further extended by formulating the forcing terms corresponding to static and time-dependent magnetic fields generated through alternative mechanisms. Also, the present formulation does not work for a bubble placed in a region of constant applied magnetic field for which a modified theory is required.

Despite these limitations, this work lays a foundational framework for studying non-spherical oscillations of encapsulated magnetic bubbles, offering a basis for further advancements in the field.\\

\justify{}
\textbf{Acknowledgement.} AKB would like to thank Shaik Nadeem Karim for useful discussions. AKB and GT acknowledge the Ministry of Education (MoE), Government of India, for providing research support.\\

\justify{}
\textbf{Funding.} This work was supported by the Science and Engineering Research Board, Department of Science and Technology, Government of India, under Project No. CRG/2022/005775.\\
\justify{
\textbf{Declaration of Interests.} The authors report no conflict of interest.\\
}
\appendix
\section{Magnetic forces coefficients for coil field}\label{A:1}
\begin{align}
\mathcal{H}_0=\frac{\mu_0\chi_sH_2^2R_0^2}{2Ra^2},\quad\mathcal{H}_2=\frac{\mu_0\chi_sH_2^2R_0^2}{2Ra^2},\quad\mathcal{H}=\frac{\mu_0\chi_sH_2^2R_0^2}{4R^2a^2},\quad\mathcal{H}^1_2=\frac{\mu_0\chi_sH_2^2R_0^2}{4Ra^2}
\end{align}
\begin{align}
\mathcal{F}_1&=\frac{-13 k^2-13k+6}{(2k-1)(2k+3)} & \mathcal{F}^1_1&=\frac{5(2k^2+2k-3)}{(2k-1)(2k+3)}\\
\mathcal{F}_2&=\frac{10k^2(k^2+2k+1)}{(2k-1)(2k+3)} & \mathcal{F}^1_2&=\frac{3(k^2+k+3)}{(2k-1)(2k+3)}\\
\mathcal{F}_3&=\frac{-3(k^3+7k^2+14k+8)}{(2k+3)(2k+5)} & \mathcal{F}^1_3&=\frac{3(k^2+6k+8)}{(2k+3)(2k+5)}\\
\mathcal{F}_4&=\frac{3(k^4+10k^3+35k^2+50k+24)}{(2k+3)(2k+5)} & \mathcal{F}^1_4&=\frac{-3k(k^3+9k^2+26k+24)}{(2k+3)(2k+5)}
\end{align}
\begin{align}
\mathcal{F}_5&=\frac{-3k(2k^3-5k^2+8k-5)}{(2k-1)(2k-3)} & \mathcal{F}^1_5&=\frac{3(k^2-4k+3)}{(2k-1)(2k-3)}\\
\mathcal{F}_6&=\frac{3k(2k^4-k^3-16k^2+25k-10)}{(2k-1)(2k-3)} & \mathcal{F}^1_6&=\frac{3(k^3-6k^2+11k-6)}{(2k-1)(2k-3)}
\end{align}
\section{Magnetic forces coefficients for dipole field}
\begin{align}
\mathcal{H}_0=\frac{24\mu_0\chi_sM^2R_0^2}{z^8R} ,\quad\mathcal{H}_2=\frac{-24\mu_0\chi_sM^2R_0^2}{z^8R},\quad\mathcal{H}=\frac{-36\mu_0\chi_sM^2R_0^2}{z^8R},\quad\mathcal{H}^1_2=\frac{-12\mu_0\chi_sM^2R_0^2}{z^8R}
\end{align}
\begin{align}
\mathcal{F}_1&=\frac{k^2-2}{4k^2+4k-3} & \mathcal{F}^1_1&=-\frac{2k^2+2k+1}{4k^2+4k-3}\\
\mathcal{F}_2&=-\frac{k(k^3+2k^2-k-3)}{4k^2+4k-3} & \mathcal{F}^1_2&=\frac{k^2+k+3}{4k^2+4k-3}\\
\mathcal{F}_3&=-\frac{k^3+7k^2+14k+8}{4k^2+16k+15} & \mathcal{F}^1_3&=\frac{2k^3+13k^2+26k+16}{8k^3+36k^2+46k+15}\\
\mathcal{F}_4&=\frac{k^4+10k^3+35k^2+50k+24}{4k^2+16k+15} & \mathcal{F}^1_4&=-\frac{2k^4+23k^3+89k^2+138k+72}{8k^3+36k^2+46k+15}\\
\mathcal{F}_5&=-\frac{k(k^2-2k+1)}{4k^2-8k+3} & \mathcal{F}^1_5&=\frac{k^2-6k+7}{4k^2-8k+3}\\
\mathcal{F}_6&=\frac{k(k^3-2k^2-k+2)}{4k^2-8k+3} & \mathcal{F}^1_6&=\frac{k^3-6k^2+15k-14}{4k^2-8k+3}
\end{align}
\bibliographystyle{jfm}
\bibliography{references}

\end{document}


\maketitle

The optimized values of interface and material parameters obtained from the optimization problem for various natural configuration outer radii of the encapsulated bubble are tabulated in tables~\ref{stb:1} and \ref{stb:2}, respectively. The tabulated data below has been used to obtain figures$~4$ and $5$ in the main text.

\begin{table}
\begin{center}
\begin{tabular}{ccccccccccc}
~ & \multicolumn{10}{c}{{\bf Interface parameters} $\gamma_{ij}\left(\times10^{-2}\right)$}\\
$R_{20}$ \quad & $\gamma_{11}$ \quad & $\gamma_{21}$ \quad & $\gamma_{12}$ \quad & $\gamma_{22}$ \quad & $\gamma_{13}$ \quad & $\gamma_{23}$ \quad &  $\gamma_{14}$ \quad & $\gamma_{24}$ \quad & $\gamma_{15}$ \quad & $\gamma_{25}$\\
0.80 \quad & 1.00 \quad & 1.00 \quad & ~1.00 \quad & 12.00 \quad & ~3.00 \quad & 1.00 \quad & 1.00 \quad & 1.00 \quad & 7.00 \quad & 1.00 \\
1.00 \quad & 1.00 \quad & 1.00 \quad & ~1.00 \quad & ~1.00 \quad & ~1.01 \quad & 7.83 \quad & 1.00 \quad & 1.00 \quad & 8.00 \quad & 1.00 \\
1.30 \quad & 1.03 \quad & 1.00 \quad & ~1.00 \quad & ~1.00 \quad & ~3.04 \quad & 5.95 \quad & 1.00 \quad & 1.00 \quad & 8.00 \quad & 1.00 \\
1.40 \quad & 1.00 \quad & 1.00 \quad & ~1.00 \quad & ~1.00 \quad & ~1.65 \quad & 4.51 \quad & 1.00 \quad & 1.00 \quad & 8.00 \quad & 1.00 \\
1.50 \quad & 1.00 \quad & 1.00 \quad & ~1.00 \quad & ~1.94 \quad & ~6.71 \quad & 2.47 \quad & 1.00 \quad & 1.00 \quad & 8.00 \quad & 1.00 \\
1.60 \quad & 1.00 \quad & 1.00 \quad & ~1.00 \quad & ~1.00 \quad & ~4.43 \quad & 4.57 \quad & 3.36 \quad & 1.00 \quad & 8.00 \quad & 1.00 \\
1.70 \quad & 1.10 \quad & 1.00 \quad & ~1.00 \quad & ~1.00 \quad & ~8.00 \quad & 4.49 \quad & 4.49 \quad & 1.00 \quad & 8.00 \quad & 1.00 \\
1.75 \quad & 1.00 \quad & 1.00 \quad & ~1.00 \quad & ~1.00 \quad & ~8.00 \quad & 1.00 \quad & 1.00 \quad & 1.00 \quad & 8.00 \quad & 1.00 \\
1.80 \quad & 1.00 \quad & 1.00 \quad & ~1.00 \quad & ~2.85 \quad & ~8.00 \quad & 1.00 \quad & 1.00 \quad & 1.00 \quad & 8.00 \quad & 1.00 \\
1.85 \quad & 1.00 \quad & 1.00 \quad & ~1.24 \quad & ~1.00 \quad & ~7.69 \quad & 1.30 \quad & 1.00 \quad & 1.30 \quad & 8.00 \quad & 1.00 \\
1.90 \quad & 1.00 \quad & 1.00 \quad & ~1.00 \quad & ~1.00 \quad & ~8.00 \quad & 1.00 \quad & 1.00 \quad & 1.00 \quad & 8.00 \quad & 1.00 \\
2.00 \quad & 1.00 \quad & 1.00 \quad & ~1.00 \quad & ~4.00 \quad & ~8.00 \quad & 1.00 \quad & 1.00 \quad & 1.00 \quad & 8.00 \quad & 1.00 \\
2.10 \quad & 1.00 \quad & 1.00 \quad & ~1.31 \quad & ~2.33 \quad & ~8.00 \quad & 1.00 \quad & 1.00 \quad & 1.00 \quad & 8.00 \quad & 1.00 \\
2.25 \quad & 1.00 \quad & 1.00 \quad & ~6.87 \quad & ~4.05 \quad & ~8.00 \quad & 1.00 \quad & 1.00 \quad & 1.00 \quad & 8.00 \quad & 1.00 \\
2.40 \quad & 1.00 \quad & 1.00 \quad & ~1.00 \quad & 12.52 \quad & ~8.00 \quad & 1.00 \quad & 1.00 \quad & 1.00 \quad & 8.00 \quad & 1.00 \\
2.50 \quad & 1.00 \quad & 1.00 \quad & ~1.00 \quad & 30.75 \quad & ~8.00 \quad & 1.00 \quad & 8.00 \quad & 1.00 \quad & 8.00 \quad & 1.00 \\
2.70 \quad & 1.00 \quad & 1.00 \quad & ~1.00 \quad & 30.81 \quad & ~8.00 \quad & 1.00 \quad & 1.00 \quad & 1.00 \quad & 8.00 \quad & 1.00 \\
3.00 \quad & 1.00 \quad & 1.00 \quad & ~1.13 \quad & 31.28 \quad & ~8.00 \quad & 1.00 \quad & 1.00 \quad & 1.00 \quad & 8.00 \quad & 1.00 \\
3.25 \quad & 1.00 \quad & 1.00 \quad & ~6.28 \quad & 48.12 \quad & ~8.00 \quad & 1.00 \quad & 1.00 \quad & 1.00 \quad & 8.00 \quad & 1.00 \\
\end{tabular}
\caption{Optimized interface parameters ({IP}) for encapsulated bubbles with different natural configuration outer radii $R_{20}\,(\mu$m), excitation pressure $p_a=0.15\,$MPa and frequency $f=2.5\,$MHz.}
\label{stb:1}
\end{center}
\end{table}

\begin{table}
\begin{center}
\begin{tabular}{cccccccc}
~ & \multicolumn{7}{c}{\bf Bulk material parameters}\\
$R_{20}$ \quad & $C_1$ \quad & $C_2$ \quad & $h$ \quad & $\eta^{\rm S}$ \quad & $p_{g_0}$ \quad & $\chi$ \quad & $k^{\rm S}$ \\
0.80 \quad & 2.60 \quad & 4.99 \quad &  7.40 \quad & 0.07 \quad & 0.19 \quad & 0.33 \quad & 1.55 \\
1.00 \quad & 4.43 \quad & 3.06 \quad &  7.40 \quad & 0.07 \quad & 0.18 \quad & 0.33 \quad & 1.55 \\
1.30 \quad & 3.00 \quad & 4.50 \quad &  7.50 \quad & 0.07 \quad & 0.18 \quad & 0.33 \quad &  1.57 \\
1.40 \quad & 4.49 \quad & 3.00 \quad &  7.50 \quad & 0.07 \quad & 0.18 \quad & 0.33 \quad & 1.57 \\
1.50 \quad & 3.46 \quad & 4.32 \quad &  7.87 \quad & 0.07 \quad & 0.19 \quad & 0.36 \quad & 1.65 \\
1.60 \quad & 3.00 \quad & 4.89 \quad &  7.50 \quad & 0.07 \quad & 0.24 \quad & 0.35 \quad & 1.57 \\
1.70 \quad & 4.73 \quad & 3.26 \quad &  7.59 \quad & 0.08 \quad & 0.22 \quad & 0.36 \quad & 1.70 \\
1.75 \quad & 3.03 \quad & 4.46 \quad &  7.50 \quad & 0.07 \quad & 0.24 \quad & 0.33 \quad & 1.57 \\
1.80 \quad & 3.00 \quad & 4.78 \quad &  7.20 \quad & 0.07 \quad & 0.25 \quad & 0.33 \quad & 1.51 \\
1.85 \quad & 3.08 \quad & 4.45 \quad &  7.20 \quad & 0.07 \quad & 0.24 \quad & 0.32 \quad & 1.56 \\
1.90 \quad & 3.02 \quad & 4.47 \quad &  7.30 \quad & 0.07 \quad & 0.24 \quad & 0.32 \quad & 1.53 \\
2.00 \quad & 3.00 \quad & 4.50 \quad &  7.40 \quad & 0.07 \quad & 0.26 \quad & 0.33 \quad & 1.55 \\
2.10 \quad & 3.00 \quad & 4.82 \quad &  7.90 \quad & 0.07 \quad & 0.28 \quad & 0.37 \quad & 1.66 \\
2.25 \quad & 3.03 \quad & 5.00 \quad &  7.49 \quad & 0.07 \quad & 0.30 \quad & 0.36 \quad & 1.57 \\
2.40 \quad & 3.91 \quad & 4.58 \quad &  7.50 \quad & 0.08 \quad & 0.30 \quad & 0.38 \quad & 1.80 \\
2.50 \quad & 4.50 \quad & 3.80 \quad &  7.50 \quad & 0.08 \quad & 0.30 \quad & 0.37 \quad & 1.80 \\
2.70 \quad & 5.00 \quad & 3.45 \quad &  7.80 \quad & 0.08 \quad & 0.31 \quad & 0.39 \quad & 1.87 \\
3.00 \quad & 4.50 \quad & 3.50 \quad &  8.10 \quad & 0.08 \quad & 0.31 \quad & 0.38 \quad & 1.94 \\
3.25 \quad & 4.50 \quad & 3.50 \quad &  8.10 \quad & 0.08 \quad & 0.32 \quad & 0.38 \quad & 1.93 \\
\end{tabular}
\caption{Optimized material parameters ({MP}) such as shell elastic constants $\left(C_1,C_2\right)\,$MPa, bubble shell thickness $(h)\,$nm, viscosity of shell $(\eta^{\rm S})\,$Pa-s,  natural configuration pressure $(p_{g_0})\,$MPa, shell elasticity modulus $(\chi)\,$N/m and shell dilatational viscosity $(k^{\rm {S}}\,)10^{-9}\,$kg/s for encapsulated bubbles with different natural configuration outer radii $(R_{20})\,\mu$m, viscosity of the liquid $\eta^{\rm L}=1\,$mPa-s, excitation pressure $p_a=0.15\,$MPa and frequency $f=2.5\,$MHz.}
\label{stb:2}
\end{center}
\end{table}